\newif\iffull      \fulltrue       
\newif\ifdraft     \draftfalse      

\documentclass[11pt,a4paper]{article}
\usepackage{t1enc,RR}
\RRdate{Janvier 2007}
\RRauthor{ Alain Frisch\thanks{INRIA, projet Gallium} \and Haruo Hosoya\thanks{University of Tokyo} }
\RRtitle{ Vers un typage praticable pour les macro transducteurs
d'arbre }
\RRetitle{ Towards Practical Typechecking for Macro Tree Transducers }
\RRabstract{Macro tree transducers (mtt) are an important model that both covers
many useful XML transformations and allows decidable exact
typechecking.  This paper reports our first step toward an
implementation of mtt typechecker that has a practical efficiency.
Our approach is to represent an input type obtained from a backward
inference as an alternating tree automaton, in a style similar to
Tozawa's XSLT0 typechecking.  In this approach, typechecking
reduces to checking emptiness of an alternating tree automaton.
We propose several optimizations (Cartesian factorization, state
partitioning) on the backward inference process in order to produce
much smaller alternating tree automata than the naive algorithm,
and we present our
efficient algorithm for checking emptiness of alternating tree
automata, where we exploit the explicit representation of alternation
for local optimizations.
Our preliminary experiments confirm that our
algorithm has a practical performance that can typecheck simple
transformations with respect to the full XHTML in a reasonable time.

}
\RRresume{Les macro transducteurs d'arbre (mtt) constituent un modèle
important, dans la mesure où ils permettent de réaliser de nombreuses
transformations XML et où ils admettent un typage exact décidable.
Cet article rend compte d'une première étape en direction de
l'implémentation d'un typeur pour les mtt efficace en pratique. Notre
approche consiste à représenter le type d'entrée obtenu par inférence
inverse sous la forme d'un automate d'arbre alternant, dans un style
similaire à celui introduit par Tozawa pour le typage de XSLT0. Le
problème de la vérification du bon typage du transducteur se réduit
alors à celui du test de vide pour un automate d'arbre alternant.
Nous
proposons plusieurs optimisations (factorisation cartésienne,
partionnement des états) pour le processus d'inférence inverse, avec
l'objectif de produire des automates alternants significativement plus
petits qu'avec l'algorithme naïf. Nous décrivons également un
algorithme efficace pour le test de vide pour un automate d'arbre
alternant, dans lequel nous exploitons la représentation explicite de
l'alternation pour permettre des optimisations locales. Nos
expériences préliminaires confirment que notre algorithme atteint des
performances suffisantes pour typer des transformations par rapport à
la DTD XHTML complète, en un temps raisonnable.

}
\URRocq
\RRtheme{\THSym}
\RRprojet{Gallium}
\RRmotcle{automates d'arbre, transducteurs d'arbre, typage exact,
automates alternants}
\RRkeyword{tree automata, tree transducers, exact typechecking,
alternating automata}

\usepackage{local}
\usepackage{fullpage}
\usepackage{stmaryrd}
\usepackage{amssymb}
\usepackage{amsmath}
\usepackage{alltt}
\usepackage{url}

\title{Towards Practical Typechecking for Macro Tree Transducers}
\author{Alain Frisch \and Haruo Hosoya}
\date{}

\newcommand{\A}{{\cal A}}

\newcommand{\M}{{\cal M}}
\newcommand{\N}{{\cal N}}
\newcommand{\T}{{\cal T}}
\newcommand{\U}{{\cal U}}

\newcommand{\lang}{{\cal L}}
\newcommand{\an}[2]{{#1}^{(#2)}}

\begin{document}
\makeRR


\section{Introduction}
\label{sec:intro}

Static typechecking for XML transformations is an important problem
that has expectedly a significant impact on real-world XML
developments.  To this end, several research groups have made efforts
in building typed XML programming languages
\cite{HosoyaPierceTOIT03,cduce2003} with much influence from the
tradition of typed functional languages
\cite{AppelMacQueenSMLNJ,ocaml}.  While this line of work has
successfully treated general, Turing-complete languages, its
approximative nature has resulted in an even trivial transformation
like the identity function to fail to typecheck unless a large amount
of code duplicates and type annotations are introduced
\cite{HosoyaFiltersJFP}.  Such situation has led us to pay attention
to completely different approaches that have no such deficiency, among
which {\em exact typechecking} has emergingly become promising.  The
exact typechecking approach has extensively been investigated for
years
\cite{Maneth07,Perst04,Milo:PODS2000,Tozawa01,Tozawa04,Tozawa06,Maneth05,Milo:PODS99,Milo:PODS01,MartensICDT2003,Murata97,Martens04},
in which {\em macro tree transducers} (mtt) have been one of the most
important models since they allow decidable exact typechecking
\cite{Engelfriet85}, yet cover many useful XML transformations
\cite{Engelfriet85,Maneth05,Engelfriet03,Nakano06}.  Unfortunately,
these studies are mainly theoretical and their practicality has never
been clear except for some small cases \cite{Tozawa01,Tozawa04}.

This paper reports our first step toward a {\em practical}
implementation of typechecker for mtts.  As a basic part, we follow an
already-established scheme called backward inference, which computes
the preimage of the output type for the subject transformation and
then checks it against the given input type.  This is because, as
known well, the more obvious, forward inference does not work since
the image of the input type is  not always a regular tree language in
general.  Our proposal is, on top of this scheme, to use a
representation of  the preimage by an {\em alternating tree automaton}
\cite{Slutzki85}, extending the idea used in Tozawa's typechecking for
XSLT0 \cite{Tozawa01}. In this approach, typechecking
reduces to checking emptiness of an alternating tree automaton. 

Whereas normal tree automata use only disjunctions in the transition
relation, alternating tree automata can use both disjunctions and
conjunctions.  This extra freedom permits a more compact
representation (they can be exponentially more succinct than normal
tree automata) and make them a good intermediate language to study
optimizations. Having explicit representation of transitions as
Boolean formulas allowed us to derive optimized versions of the rules
for backward inference, such as Cartesian decomposition or state
partitioning (Section~\ref{sec:backwardalgo}). These optimizations
allow our algorithm to scale to large types.  We also use Boolean
reasoning to derive an efficient emptiness algorithm for alternating
tree automata (Section~\ref{sec:emptiness}). For instance, this
algorithm uses the following fact as an efficient shortcut: when
considering a formula $\phi = \phi_1 \wedge \phi_2$, if $\phi_1$ turns
out to denote an empty set, then so is $\phi$, and thus the algorithm
doesn't even need to look at $\phi_2$.  Note that the exploited fact
is immediately available in alternating tree automata, while it is not
in normal tree automata.

We have made extensive experiments on our implementation.  We have
written several sizes of transformations and verified against the full
XHTML automatically generated from its DTD (in reality,
transformations are often small, but types that they work on are quite
big in many cases; excellent statistical evidences are provided in
\cite{Moeller2005}.)  The results show that, for this scale of
transformations, our implementation has successfully completed
typechecking in a reasonable time even with XHTML, which is considered
to be quite large.
We have also compared the performance of our implementation with
Tozawa and Hagiya's \cite{Tozawa04} and confirmed that ours has
comparable speed for their small examples that are used in their own
experiments.

On the theoretical side, we have established an exact relationship
with two major existing algorithms for mtt typechecking, a classical
algorithm based on ``function enumeration'' \cite{Engelfriet03} and an
algorithm proposed by Maneth, Perst, and Seidl (MPS algorithm)
\cite{Maneth07}.  Concretely, we have proved that (1) the classical
algorithm  is identical to our algorithm followed by determinization
of an alternating tree automaton, and that (2) MPS algorithm is
identical to our algorithm followed by emptiness test of an
alternating tree automaton.  A particular implication is that our
algorithm inherits one of useful properties of MPS algorithm:
polynomial-time complexity under the restriction of  a bounded number
of copying \cite{Maneth07} (mtt typechecking is in general
exponential-time complete).  The proofs appear in the appendix,
however, since this paper is focused rather on the practical side.

\paragraph{Related work}

Numerous techniques for exact typechecking for XML transformations have
been proposed.  Many of these take their target languages from the
tree transducer family.  Those include techniques for macro tree
transducers \cite{Maneth07,Engelfriet03}, for macro forest transducers
\cite{Perst04}, for $k$-pebble tree transducers
\cite{Milo:PODS2000,Engelfriet03}, for subsets of XSLT
\cite{Tozawa01,Tozawa04}, for high-level tree transducers
\cite{Tozawa06}, and a tree transformation language TL \cite{Maneth05}.
Other techniques treat XML query languages in the select-construct
style \cite{Milo:PODS99,Milo:PODS01,MartensICDT2003} or even simpler
transformations \cite{Murata97,Martens04}.  Most of the above
mentioned work provides only theoretical results; the only exceptions
are \cite{Tozawa01,Tozawa04}, where some experimental results are
shown though we have examined much bigger examples (in particular in
the size of types).

Several algorithms in pragmatic approaches have been proposed to
address high complexity problems related to XML typechecking.  A
top-down algorithm for inclusion test on tree automata has been
developed and used in XDuce typechecker \cite{HVPRegtypeToplas}; an
improved version is proposed in \cite{SudaHosoya05}.  A similar idea
has been exploited in the work on {$\mathbb{C}$Duce} on the emptiness
check for alternating tree automata \cite{Frisch04PhD}; the emptiness
check algorithm in our present work is strongly influenced by this.
Tozawa and Hagiya have developed BDD-based algorithms for inclusion
test on tree automata \cite{TozawaHagiya03} and for satisfiability
test on a certain logic related to XML typechecking \cite{Tozawa04}.


\paragraph{Overview}

This paper is organized as follows. In Section~\ref{sec:prelim}, we
recall the classical definitions of macro tree transducers (mtt),
bottom-up tree automata (bta), and alternating tree automata (ata).
In Section~\ref{sec:typecheck}, we present the two components of our
typechecking algorithm: backward type inference (which produces an ata
from an mtt and a deterministic bta) and emptiness check for
alternating tree automata. In Section~\ref{sec:algo}, we revisit these
two components from a practical point of view and we describe
important optimizations and implementation techniques. In
Section~\ref{sec:exper}, we report the results of our experiments with
our implementation of the typechecker for several XML transformations.
In Section~\ref{sec:concl}, we conclude this paper with our future
direction.  Appendix~\ref{sec:comparison} is devoted to a precise
comparison between our algorithm and the classical algorithm or the
Maneth-Perst-Seidl algorithm for typechecking mtt. We show that each
of these algorithms can be retrieved from ours by composing with a
know algorithm. In Appendix~\ref{sec:bounded}, we propose the notion
of bounded-traversing alternating tree automata, which is a natural
counterpart of syntactical bounded-copying mtts as proposed in
\cite{Maneth07}. We show in particular that this notion ensures that
the emptiness check runs in polynomial time.


\section{Preliminaries}
\label{sec:prelim}

\subsection{Macro Tree Transducers}
\label{sec:mtt}

We assume an alphabet $\Sigma$ where each symbol $a\in \Sigma$ is
associated with its arity; often we write $\an{a}{n}$ to denote a symbol
$a$ with arity $n$.  We assume that there is a symbol $\eps$ with
zero-arity.  {\em Trees}, ranged over by $v,w,\ldots$, are defined as
follows:
\iffull
$$
\begin{array}{lll}
v & ::= & a^{(n)}(v_1,\ldots,v_n)
\end{array}
$$
\fi
We write $\eps$ for $\eps()$ and $\vec{v} = (v_1,\ldots,v_n)$ 
to represent a tuple of trees.
\iffull \fi %
Assume a set of variables, ranged over by $x,y,\ldots$.  A {\em macro
  tree transducer} (mtt) $\T$ is a tuple $(P, P_0, \Pi)$ where $P$ is a
finite set of procedures, $P_0\subseteq P$ is a set of initial
procedures, and $\Pi$ is a set of (transformation) rules each of the
form
\iffull
$$
\an{p}{k}(\an{a}{n}(x_1, \ldots, x_n), y_1,\ldots,y_k) \to e
$$
\fi
where each $y_i$ is called {\em (accumulating) parameter} and $e$
is a $(n,k)$-expression. We will abbreviate the tuples
$(x_1,\ldots,x_n)$ and $(y_1,\ldots,y_k)$ to $\vec{x}$ and  $\vec{y}$.
Note that each procedure is associated with its arity, i.e., the
number of parameters; we write $p^{(k)}$ to denote a procedure $p$
with arity $k$.  An $(n,k)$-expression $e$ is defined by the following
grammar
$$
\begin{array}{lll}
e & ::= & \an{a}{m}(e_1,\ldots,e_m) \mid \an{p}{l}(x_h, e_1,\ldots,e_l) \mid y_j
\end{array}
$$
where only $y_j$ with $1\leq j \leq k$ 
and $x_h$ with $1 \leq h \leq n$ can appear as variables.
We assume that each initial procedure has arity zero.

We describe the semantics of an mtt $(P, P_0, \Pi)$ by a denotation
function $\den{\cdot}$.  First, the semantics of a procedure
$\an{p}{k}$ takes a tree $\an{a}{n}(v_1, \ldots, v_n)$ and parameters
$\vec{w} = (w_1,\ldots,w_k)$ and returns the set of trees resulted
from evaluating any of $p$'s body expressions.
$$
\den{\an{p}{k}}(\an{a}{n}(\vec{v}), \vec{w}) =
\bigcup_{(\an{p}{k}(\an{a}{n}(\vec{x}), \vec{y}) \to e) \in \Pi}
\den{e}(\vec{v}, \vec{w})
$$
Then, the semantics of an $(n,k)$-expression $e$ takes a current
$n$-tuple $\vec{v} = (v_1,\ldots,v_n)$ of trees and a $k$-tuple of
parameters $\vec{w} = (w_1,\ldots,w_k)$, and returns the set of trees
resulted from the evaluation. It is defined as follows.
$$
\begin{array}{lll}
\den{\an{a}{m}(e_1, \ldots, e_m)}(\vec{v}, \vec{w}) & = &
\{ \an{a}{m}(v'_1, \ldots, v'_m) \mid
 v'_i \in \den{e_i}(\vec{v}, \vec{w}) \mbox{ for } i=1,\ldots,m
\}
\\
\den{\an{p}{l}(x_h, e_1,\ldots,e_l)}(\vec{v}, 
\vec{w}) & = &
\{ \den{\an{p}{l}}(v_h, (w_1',\ldots,w_l')) \mid
w_j' \in \den{e_j}(\vec{v}, \vec{w}) 
\mbox{ for } j=1,\ldots,l \}
\\
\den{y_j}(\vec{v}, \vec{w}) & = & \{ w_j \}
\end{array}
$$
\iffull
A constructor expression $\an{a}{m}(e_1, \ldots, e_m)$ evaluates
each subexpression $e_i$ and reconstructs a tree node with $a$ and the
results of these subexpressions.   A procedure call $p(x_h,
e_1,\ldots,e_l)$ evaluates the procedure $p$ under the $h$-th subtree
$v_h$, passing the results of $e_1,\ldots,e_l$ as parameters.
A variable expression $y_j$ simply results in the corresponding
parameter's value $w_j$.  
\fi
Note that an mtt is allowed to inspect only
the input tree and never a part of the output tree being constructed.
Also, parameters only accumulate subtrees that will potentially become
part of the output and never point to parts of the input.   

The whole semantics of the mtt with respect to a given input tree $v$
is defined by $\T(v) = \bigcup_{p_0\in P_0}\den{p_0}(v)$.  An mtt $\T$
is deterministic when $\T(v)$ has at most one element for any $v$;
also, $\T$ is total when $\T(v)$ has at least one element for any $v$.
We will also use the classical definition of images and preimages:
$\T(V) = \bigcup_{v \in V} \T(v)$, $\T^{-1}(V') = \{v ~|~ \exists v'
\in V'. v' \in \T(v) \}$.

\subsection{Tree Automata and Alternation}
\label{sec:ta}

\newcommand{\from}{\leftarrow}
\newcommand{\down}{\downarrow}

A {\em (bottom-up) tree automaton} (bta) $\M$ is a tuple $(Q, Q_F,
\Delta)$ where $Q$ is a finite set of states, $Q_F\subseteq Q$ is a
set of final states, and $\Delta$ is a set of (transition) rules each
of the form $q \from \an{a}{n}(q_1,\ldots,q_n)$ where each $q_i$ is
from $Q$. We will write $\vec{q}$ for the tuple
$(q_1,\ldots,q_n)$. Given a bta $\M = (Q, Q_F, \Delta)$, acceptance of
a tree by a state is defined inductively as follows: $\M$ accepts a
tree $\an{a}{n}(\vec{v})$ by a state $q$ when there is a rule $q \from
\an{a}{n}(\vec{q})$ in $\Delta$ such that each subtree $v_i$ is
accepted by the corresponding state $q_i$.  $\M$ accepts a tree $v$
when $\M$ accepts $v$ by a final state $q\in Q_F$. We write
$\den{q}_\M$ for the set of trees that the automaton $\M$ accepts by
the state $q$ (we drop the subscript $\M$ when it is clear), and
$\lang(\M) = \bigcup_{q \in Q_F} \den{q}$ for  the set of trees
accepted by the automaton $\M$.  Also, we sometimes say that a value
$v$ {\em has type} $q$ when $v$ is accepted by the state $q$.
A bta $(Q, Q_F, \Delta)$ is complete and deterministic when, for any
constructor $\an{a}{n}$ and $n$-tuple of states $\vec{q}$,
there is exactly one transition
rule of the form $q \from \an{a}{n}(\vec{q})$ in $\Delta$.
Such a bta is called {\em deterministic bottom-up tree automaton}
(dbta).  For any value $v$, there is exactly one state $q$ such that
$v \in \den{q}$. In other words,  the collection $\{ \den{q} ~|~ q \in Q \}$
is a partition of the set of trees.   

An {\em alternating tree automaton} (ata) $\A$ is a tuple $(\Xi,
\Xi_0, \Phi)$ where $\Xi$ is a finite set of states, $\Xi_0\subseteq
\Xi$ is a set of initial state, and $\Phi$ is a function that maps each
pair $(X,\an{a}{n})$ of a state and an $n$-ary constructor
to an $n$-formula, where $n$-formulas are defined by the following
grammar.
$$
\begin{array}{lll}
\phi & ::= & \down_i X \mid \phi_1\vee\phi_2 
\mid \phi_1\wedge\phi_2 \mid \top \mid \bot
\end{array}
$$
(with $1 \leq i \leq n$). In particular, note that a $0$-ary
formula evaluates naturally to a Boolean.  Given an ata $\A = (\Xi,
\Xi_0, \Phi)$, we define acceptance of a tree by a state.  $\A$ accepts
a tree $\an{a}{n}(\vec{v})$ by a state $X$ when $\vec{v} \p
\Phi(X,\an{a}{n})$ holds, where the judgment $\vec{v} \p \phi$ is
defined inductively as follows: \iffull
\begin{itemize}
\item $\vec{v} \p \phi_1 \wedge \phi_2$ if $\vec{v} \p \phi_1$ and $\vec{v} \p \phi_2$.
\item $\vec{v} \p \phi_1 \vee \phi_2$ if $\vec{v} \p \phi_1$ or $\vec{v} \p \phi_2$.
\item $\vec{v} \p \top$.
\item $\vec{v} \p \down_i X$ if $\A$ accepts $v_i$ by $X$.
\end{itemize}
\fi
That is, $\vec{v} \p \phi$ intuitively means that $\phi$ holds by
interpreting each $\down_i X$ as ``$v_i$ has type $X$.''  We write
$\den{X}$ for the set of trees accepted by a state $X$ and $\den{\phi}
= \{ \vec{v} ~|~ \vec{v} \p \phi \}$ for the set of $n$-tuples
accepted by an $n$-formula $\phi$. We write $\lang(\A) = \bigcup_{X_0
  \in \Xi_0} \den{X_0}$ for the language accepted by the ata $\A$.
Note that a bta $\M = (Q,Q_F,\Delta)$ can be seen as an ata with the
same set of states and final states by defining the function $\Phi$ as
$\Phi(q,\an{a}{n}) = \bigvee_{(q \from \an{a}{n}(\vec{q})) \in \Delta}
\bigwedge_{i=1,..,n} \down_i q_i$, and the definitions for the
semantics of states and the language accepted by the automaton seen as
a bta or an ata then coincide.  We will use the notation
$\simeq$ to represent semantical equivalence of pairs of states or
pairs of formulas.

\section{Typechecking}
\label{sec:typecheck}

\newcommand{\Inf}{\mathrm{Inf}}
\newcommand{\DInf}{\mathrm{DInf}}
\newcommand{\Spec}{\mathrm{Spec}}
\newcommand{\IN}{\mathrm{in}}
\newcommand{\OUT}{\mathrm{out}}

\subsection{Backward inference}
\label{sec:backward-formal}

Given a dbta $\M_\OUT$ (``output type''), a bta $\M_\IN$ (``input
type''), and an mtt $\T$, the goal of typechecking is to verify that
$\T(\lang(\M_\IN)) \subseteq \lang(\M_\OUT)$.  It is well known that
$\T(\lang(\M_\IN))$ is in general beyond regular tree languages and
hence the forward inference approach (i.e., first calculate an
automaton representing $\T(\lang(\M_\IN))$ and check it to be included
in $\lang(\M_\OUT)$) does not work.  Therefore an approach usually
taken is the backward inference, which is based on the observation
that $\T(\lang(\M_\IN)) \subseteq \lang(\M_\OUT) \iff \lang(\M_\IN)
\cap \T^{-1}(\lang(\M)) = \emp$, where $\M$ is the complement
automaton of $\M_\OUT$.  Intuitively, if the intersection
$\lang(\M_\IN) \cap \T^{-1}(\lang(\M))$ is not empty, then it is
possible to exhibit a tree $v$ in this intersection.  Since this tree
satisfies that $v \in \lang(\M_\IN)$ and $\T(v) \not \subseteq
\lang(\M_\OUT)$, it means that there is a counter-example of the
well-typedness of the mtt with respect to the given input and output
types.  Algorithmically, the approach consists of computing an
automaton $\A$ representing $\T^{-1}(\lang(\M))$ and then checking
that $\lang(\M_\IN)\cap \lang(\A)=\emp$.  Since the language
$\T^{-1}(\lang(\M))$ is regular and indeed such automata $\A$ can
effectively be computed, the above disjointness is decidable.

The originality of our approach is to compute $\A$ as an alternating
tree automaton. Let a dbta $\M = (Q, Q_F, \Delta)$ and an mtt $\T =
(P, P_0, \Pi)$ be given.  Here, note that the automaton $\M$, which
denotes the complement of the output type $\M_\OUT$, can be obtained
from $\M_\OUT$ in a linear time since $\M_\OUT$ is deterministic.
From $\M$ and $\T$, we build an ata $\A = (\Xi, \Xi_0, \Phi)$ where
$$
\begin{array}{lll}
\Xi & = &
\{ \tup{\an{p}{k},q,\vec{q}} \mid \an{p}{k}\in P, \; 
q \in Q, \vec{q} \in Q^k \}
\\
\Xi_0 & = & \{ \tup{p_0, q} \mid p_0\in P_0,\; q\in Q_F \}
\\
\Phi(\tup{\an{p}{k}, q, \vec{q}},\an{a}{n}) & = & 
\displaystyle
\bigvee_{(\an{p}{k}(\an{a}{n}(\vec{x}), \vec{y})\to e) \in \Pi} 
\Inf(e, q, \vec{q}).
\end{array}
$$
Here, the function $\Inf$ is defined inductively as follows.
$$
\begin{array}{lll}
\Inf(\an{b}{m}(e_1, \ldots, e_m), q, \vec{q}) & = &
\displaystyle
\bigvee_{(q \from \an{b}{m}(\vec{q'}))\in \Delta}
\bigwedge_{j=1,..m} \Inf(e_j, q_j', \vec{q})
\\
\Inf(\an{p}{l}(x_h, e_1, \ldots, e_l), q, \vec{q}) & = &
  \displaystyle
  \bigvee_{\vec{q'} \in Q^l}
  \left(\down_h \tup{\an{p}{l}, q, \vec{q'}} \wedge
    \bigwedge_{j=1,..,l} \Inf(e_j, q_j', \vec{q})
   \right)
\\
\Inf(y_j, q, \vec{q}) & = &
\left\{
  \begin{array}{ll}
    \top & \hfill (q = q_j) \\
    \bot & \hfill (q \neq q_j) 
  \end{array}
\right.
\end{array}
$$

\iffull

Let us explain why this algorithm works.  Since a precise
discussion is critical for understanding subsequent sections, we
summarize our justification here as a formal proof.

\begin{theorem}
  $\lang(\A) = \T^{-1}(\lang(\M))$.
\end{theorem}

\begin{pf}
  Intuitively, each state $\tup{p,q,\vec{q}}$ represents the
  set of trees $v$ such that the procedure $p$ may transform  $v$ to
  some tree $u$ of type $q$, assuming that the parameters $y_i$
  are bound to trees $w_i$ each of type $q_i$. Formally, we
  prove the following invariant
  \begin{equation}
    \label{eq:inv-proc}  
    \forall v.~ \forall \vec{w} \in \den{\vec{q}}.~
    v\in \den{\tup{\an{p}{k},q,\vec{q}}} \iff
    \den{\an{p}{k}}(v, \vec{w})\cap \den{q} \neq\emp 
  \end{equation}
  where $\vec{w} \in \den{\vec{q}}$ means $w_1 \in \den{q_1}, \ldots,
  w_k \in \den{q_k}$.
  Note that this invariant implies that the right-hand side does not
  depend on the specific choice of the values $w_i$ from the sets
  $\den{q_i}$; this point will be crucial later.  From this invariant,
  the initial states $\Xi_0$ represent the set of trees that we want
  and hence the result follows:
  $$
  \begin{array}{lll}
    \lang(\A) &=&
    \bigcup\{ \den{\tup{p_0,q}} \mid p_0\in P_0,\; q\in Q_F \}\\
    &=&
    \{ v \mid \den{p_0}(v) \cap \den{q} \neq \emp, \; p_0\in P_0,\; q\in
    Q_F \}\\
    &=&
    \{ v \mid \T(v) \cap \lang(\M) \neq \emp \} \\
    &=&
    \T^{-1}(\lang(\M))
  \end{array}
  $$
  The proof of the invariant (\ref{eq:inv-proc}) proceeds by induction
  on the structure of $v$.  For the proof, we first need to consider
  an invariant that holds for the function $\Inf$.  Informally,
  $\Inf(e, q, \vec{q})$ infers an $n$-formula representing the set of
  $n$-tuples $\vec{v}$ such that the expression $e$ may transform
  $\vec{v}$ to some tree of type $q$, assuming that the parameters
  $y_i$ are bound to trees $w_i$ each of type $q_i$.  Formally, we
  prove the following:
  \begin{equation}
    \label{eq:inv-expr}
    \forall \vec{v}.~ \forall \vec{w} \in \den{\vec{q}}.~
    \vec{v} \in \den{\Inf(e,q,\vec{q})} \iff
    \den{e}(\vec{v}, \vec{w})\cap \den{q} \neq\emp 
  \end{equation}
  Indeed, this implies the invariant (\ref{eq:inv-proc}).  
  Let $v=\an{a}{n}(\vec{v})$; for all $\vec{w}\in\den{\vec{q}}$:
  \begin{eqnarray*}
    v \in \den{\tup{\an{p}{k},q,\vec{q}}} & \iff &
    \vec{v} \in \den{\Phi(\tup{\an{p}{k},q,\vec{q}}, \an{a}{n})} \\
    & \iff &
    \exists (\an{p}{k}(\an{a}{n}(\vec{x}),\vec{y}) \to e) \in \Pi.~
    \vec{v} \in 
    \den{\Inf(e, q, \vec{q})} 
    \\
    & \stackrel{by (\ref{eq:inv-expr})}{\iff} &
    \exists (\an{p}{k}(\an{a}{n}(\vec{x}),\vec{y}) \to e) \in \Pi.~
    \den{e}(\vec{v}, \vec{w})\cap \den{q} \neq\emp 
    \\
    & \iff &
    \den{p}(v,\vec{w})\cap \den{q} \neq \emp
  \end{eqnarray*}
  The invariant (\ref{eq:inv-expr}) is in turn proved by
  induction on the structure of $e$.  
  \begin{description}
  \item[Case $e=\an{b}{m}(e_1, \ldots, e_m)$.]  In order for a tree
    $u$ of type $q$ to be produced from the constructor expression,
    first, there must be a transition $q \from
    \an{b}{m}(\vec{q'})\in\Delta$.  In addition, $u$'s each subtree
    must have type $q_i'$ and must be produced from the corresponding
    subexpression $e_i$.  For the latter condition, we can use the
    \ih\ for (\ref{eq:inv-expr}).  Formally, for all
    $\vec{w}\in\den{\vec{q}}$:
    \begin{eqnarray*}
      \vec{v} \in \den{\Inf(e,q,\vec{q})} 
      & \iff &
      \vec{v} \in
      \den{
        \bigvee_{q \to \an{b}{m}(\vec{q'})\in\Delta}
        \bigwedge_{j=1,\ldots,m}
        \Inf(e_j,q_j',\vec{q})} \\
      & \iff &
      \exists (q \from \an{b}{m}(\vec{q'}))\in\Delta.~ 
      \forall j=1,\ldots,m.~
      \vec{v} \in \den{\Inf(e_j,q_j',\vec{q})}
      \\
      & \stackrel{\mathrm{by I.H. for (\ref{eq:inv-expr})}}{\iff} &
      \exists (q \from \an{b}{m}(\vec{q'}))\in\Delta.~ 
      \forall j=1,\ldots,m.~
      \den{e_j}(\vec{v}, \vec{w}) \cap \den{q_j'} \neq \emp 
      \\
      & \iff &
      \den{e}(\vec{v}, \vec{w}) \cap \den{q} \neq \emp 
    \end{eqnarray*}
  \item[Case $e = \an{p}{l}(x_h, e_1, \ldots, e_l)$.]  In order for a tree $u$
    of type $q$ to be produced from the procedure call, first, a tree
    $w_j'$ of some type $q_j'$ must be yielded from each parameter
    expression $e_j$.  In addition, the $h$-th input tree must have
    type $\tup{\an{p}{l},q,(q_1',\ldots,q_l')}$ since the result tree $u$ must
    be produced by the procedure $\an{p}{l}$ from the $h$-th input tree with
    parameters $w_1',\ldots,w_l'$ of types $q_1',\ldots,q_l'$.  We can
    use the \ih\ for (\ref{eq:inv-expr}) for the former condition and
    that for (\ref{eq:inv-proc}) for the latter condition.  Formally,
    for all $\vec{w}\in\den{\vec{q}}$:
    \begin{eqnarray}
      \vec{v} \in \den{\Inf(e,q,\vec{q})} 
      & \iff &
      \vec{v} \in 
      \den{
        \bigvee_{\vec{q'}\in Q^l}
        \down_h\tup{p,q,\vec{q'}} 
        \wedge
        \bigwedge_{j=1,\ldots,l}
        \Inf(e_j, q_j', \vec{q}) }
      \nonumber
      \\
      & \iff &
      \exists \vec{q'}\in Q^l.~
      v_h \in \den{\tup{p,q,\vec{q'}}} 
      \wedge
      \forall j=1,\ldots,l.~
      \vec{v} \in 
      \den{ \Inf(e_j, q_j', \vec{q}) }
      \nonumber
      \\
      & \stackrel{\mathrm{by I.H. for (\ref{eq:inv-expr})}}{\iff} &
      \exists \vec{q'}\in Q^l.~
      v_h \in \den{\tup{p,q,\vec{q'}}} 
      \wedge
      \forall j=1,\ldots,l.~
      \den{e_j}(\vec{v},\vec{w})\cap \den{q_j'} \neq \emp
      \nonumber
      \\
      & \iff &
      \exists \vec{q'}\in Q^l.~
      \begin{array}[t]{l}
        v_h \in \den{\tup{p,q,\vec{q'}}} 
        \\
        ~\wedge~
        \exists \vec{w'}.~
        \forall j=1,\ldots,l.~
        w_j' \in \den{e_j}(\vec{v},\vec{w}) ~\wedge~
        w_j' \in \den{q_j'} 
        \label{eq:case-proccall0}
      \end{array}
    \end{eqnarray}
    We can show that the last condition holds iff
    \begin{equation}
      \exists \vec{w'}.~
      \begin{array}[t]{l}
        \den{\an{p}{l}}(v_h,\vec{w'}) \cap \den{q}\neq\emp
        ~\wedge~
        \forall j=1,\ldots,l.~
        w_j' \in \den{e_j}(\vec{v},\vec{w}) 
        \label{eq:case-proccall2}
      \end{array}
    \end{equation}
    which is equivalent to  $\den{p(x_h,e_1,\ldots,e_m)}\cap \den{q}
    \neq \emp$.  Indeed, for the ``only if'' direction, we apply the
    \ih\ for (\ref{eq:inv-proc}) where we instantiate $\vec{w}$ with
    the specific $\vec{w'}$ in  (\ref{eq:case-proccall0})---this is
    exactly the place that uses the fact that the quantification on
    $\vec{w}$ appears outside the ``$\iff$'' in
    (\ref{eq:inv-proc})---and obtain the following:
    \begin{equation}
      \exists \vec{q'}\in Q^l.~
      \exists \vec{w'}.~
      \begin{array}[t]{l}
        \den{\an{p}{l}}(v_h,\vec{w'}) \cap \den{q}\neq\emp
        \\
        ~\wedge~
        \forall j=1,\ldots,l.~
        w_j' \in \den{e_j}(\vec{v},\vec{w}) ~\wedge~
        w_j' \in \den{q_j'} 
        \label{eq:case-proccall1}
      \end{array}
    \end{equation}
    By dropping the condition $w_j' \in \den{q_j'}$ (and the unused
    quantification on $\vec{q'}$), we obtain (\ref{eq:case-proccall2}).

    For the ``if'' direction, since that the automaton $\M$
    is complete, i.e., there is in general a state $q$ for any value $w$ such
    that $w \in \den{q}$, we obtain (\ref{eq:case-proccall1}) from
    (\ref{eq:case-proccall2}).  Then, the \ih\ for (\ref{eq:inv-proc})
    yields (\ref{eq:case-proccall0}).
  \item[Case $e=y_j$.]  In order for a tree of type $q$ to be produced
    from the variable expression, $y_j$ must have type $q$.  Formally,
    first note that  $\vec{v} \in \den{\Inf(e,q,\vec{q})}  \iff  q =
    q_j$, for any $\vec{v}$.  Note also that, since $\M$ is
    deterministic bottom-up, all the states are pair-wise disjoint:
    $\den{q}\cap \den{q'}=\emp$ whenever $q\neq q'$.  Therefore, for
    all $\vec{w}\in\den{\vec{q}}$:
    \begin{eqnarray*}
      \vec{v} \in \den{\Inf(e,q,\vec{q})}
      & \iff &
      q = q_j \\
      & \iff &
      w_j\in \den{q} \\
      & \iff & 
      \den{e}(\vec{v},\vec{w}) \cap \den{q} \neq \emp
    \end{eqnarray*}
    \endofpf
  \end{description}
\end{pf}

\fi

\iffull
In the proof above, the case for variable expressions critically uses
the determinism constraint.   
Indeed, the statement of the theorem
does not necessarily hold if $\M$ is nondeterministic.  For example,
consider the nondeterministic bta $\M$ with the transition rules
$$
\begin{array}{lll}
q_0 \from b(q_1,q_2) & q_1 \from \eps & q_2 \from \eps \\
\end{array}
$$
($q_0$ is the initial state) and typecheck the mtt $\T$ with
the transformation rules
$$
\begin{array}{lll}
p_0(a(x_1)) & \to & p(x_1,\eps) \\
p(\eps, y_1) & \to & b(y_1,y_1) 
\end{array}
$$
($p_0$ is the initial procedure) with respect to the result type
$q_0$.  With this mtt, the input value $a(\eps)$ translates to
$b(\eps,\eps)$, which is accepted by $\M$.  However, our algorithm
will infer an input type that denotes the empty set, which is
incorrect.  To see this more closely, consider inference on the body
of $p$ with the result type $q=q_0$ and the parameter type $\vec{q} =
(q_1)$.  The condition (\ref{eq:inv-expr}) does not hold since the
only choice of $\vec{w}\in \den{\vec{q}}$ is $\vec{w} = (\eps)$ and,
in this case, the right hand side holds whereas the left hand side
does not since $\Inf(b(y_1,y_1), q_0, (q_1)) =
\Inf(y_1,q_1,(q_1))\wedge \Inf(y_1,q_2,(q_1)) = \top\wedge\bot =
\bot$.  The same argument can be done with the parameter type
$\vec{q}=(q_2)$.  Now, in inference on the body of $p_0$ with the
result type $q_0$, the call to $p$ must have parameter type $q_1$ or
$q_2$ since only these can accept $\eps$.  From the previous inference, we
conclude that the input type inferred for the call is again the empty
set type; so is the whole input type.

However, the variable case is the only that uses determinism.
Therefore, if the mtt uses no parameter, i.e., is a simple, top-down
tree transducer, then the same algorithm works for a non-deterministic
output type.\footnote{Completeness of the output type is not needed
  for our algorithm to work on top-down tree transducers.  This is
  because the only place where we use completeness in the proof is the
  case for procedure calls, in which completeness is actually not
  necessary if there is no parameter.}   Moreover, if the mtt $\T$ is
deterministic and total, we have $\T^{-1}(\lang(\compl{\M_\OUT})) =
\compl{\T^{-1}(\lang(\M_\OUT))}$.  It suffices to check $\lang(\M_\IN)
\subseteq \T^{-1}(\lang(\M_\OUT))$ instead of $\lang(\M_\IN) \cap
\T^{-1}(\lang(\compl{\M_\OUT})) = \emp$.   This could be advantageous
since a direct conversion from an XML schema yields a
non-deterministic automaton, and determinizing it has a potential
blow-up (though this step is known to take only a reasonable time in
practice) whereas  inclusion can be tested more efficiently  by using
known clever algorithms that avoid a full materialization of a
deterministic automaton
\cite{HVPRegtypeToplas,SudaHosoya05,TozawaHagiya03}.  Tozawa presents
in his work \cite{Tozawa01} a backward inference algorithm based on
alternating tree automata for deterministic forest transducers with no
parameters where he exploits the above observation to obtain a simple
algorithm.  \fi

\newcommand{\DNF}{\mathrm{DNF}}
\newcommand{\Xbar}{\overline{X}}
\newcommand{\Ybar}{\overline{Y}}

Finally, it remains to check $\lang(\M_\IN)\cap \lang(A)=\emp$, for
which we first calculate an ata $\A'$ representing $\lang(\M_\IN)\cap
\lang(A)$ (this can easily be done since an ata can freely use
intersections) and then check the emptiness of $\A'$. The next section
explains how to do this. The size of the ata $\A$ is polynomial in the
sizes of $\M_\OUT$ and of $\T$. The size of $\A'$ is thus polynomial
in the sizes of $\M_\IN$, $\M_\OUT$, and $\T$.

\subsection{Emptiness check}
\label{sec:emptiness-formal}

Let $\A = (\Xi,\Xi_0,\Phi)$ an alternating tree automaton. We want
to decide whether the set $\lang(\A)$ is empty or not.
We first define the following system of implications
$\rho$ where we introduce propositional variables $\Xbar$ consisting of all
subsets of $\Xi$:
$$
\begin{array}{lcl}
  \rho & = & 
  \{ 
  \Xbar \Leftarrow \Xbar_1 \wedge \ldots\wedge \Xbar_n
  \mid
  \begin{array}[t]{l}
    \exists \an{a}{n}.\;
    (\Xbar_1,\ldots,\Xbar_n)\in \DNF(\bigwedge_{X\in \Xbar}\Phi(X,\an{a}{n}))
    \}\}
  \end{array}
\end{array}
$$
Here, $\DNF(\phi)$ computes $\phi$'s disjunctive normal form  by
pushing intersections under unions and regrouping atoms of the form
$\down_i X$ for a fixed $i$; the result is formatted as a set of
$n$-tuples of state sets.  More precisely:
$$
\begin{array}{lcl}
\DNF(\top) & = & \{ (\emp,\ldots,\emp) \} \\
\DNF(\bot) & = & \emp \\
\DNF(\phi_1\wedge\phi_2) & = & 
\{ (\Xbar_1\cup \Ybar_1,\ldots,\Xbar_n\cup \Ybar_n) \mid (\Xbar_1,\ldots,\Xbar_n) \in \DNF(\phi_1),~ 
(\Ybar_1,\ldots,\Ybar_n) \in\DNF(\phi_2) \} \\
\DNF(\phi_1\vee\phi_2) & = & \DNF(\phi_1) \cup \DNF(\phi_2) \\
\DNF(\down_h X) & = & \{ (\emp,\ldots,\emp,\{ X \}, \emp,\ldots,\emp) \}
\hfill (\mbox{the $h$-th element is $\{X\}$})
\end{array}
$$
Then, with the system of implications above, we verify that $\rho\p
\{X\}$ for some $X\in \Xi_0$.  The judgment $\rho\p \Xbar$ here is defined
such that it holds when it can be derived by the single rule:
if $\rho$ contains $\Xbar\Leftarrow \Xbar_1\wedge\ldots\wedge \Xbar_n$ and 
$\rho\p \Xbar_i$ for any $i=1,\ldots,n$, then $\rho\p \Xbar$.

Each propositional variable $\Xbar$ intuitively denotes that the
intersection of the sets denoted by all the states in $\Xbar$ is
non-empty: $\bigcap_{X\in \Xbar} \den{X} \neq \emp$.  Thus, we can
prove the following.

\begin{proposition}
  $\lang(\A)\neq\emp$ iff $\rho\p \{X\}$ for some $X\in \Xi_0$.
\end{proposition}

\iffull
\begin{pf}
  The result follows by showing that $v\in \bigcap_{X\in \Xbar}\den{X}$ for some
  $v$ iff $\rho\p \Xbar$.  The ``only if'' direction can be proved by
  induction on the structure of $v$.  The ``if'' direction can be proved by
  induction on the derivation of $\rho\p \Xbar$.
\end{pf}
\fi

This emptiness check can be implemented in linear size with respect
to the size of $\rho$, which itself is exponential in the size of
$\A$.


\section{Algorithm and optimizations}
\label{sec:algo}

\newcommand{\qbar}{{\overline{q}}}
\newcommand{\cart}{\mathrm{Cart}}
\newcommand{\Choice}{\mathrm{C}}

As we explained above, our algorithm splits the type-checking process
in two phases: first, we compute an alternating tree automaton from
the output type and the mtt; second, we check emptiness of this
tree automaton. In this section, we are going to describe some details
and optimizations about these two phases.

\subsection{Backward inference}
\label{sec:backwardalgo}

A simple algorithm to compute the input type as an alternating tree
automaton is to follow naively the formal construction given in
Section~\ref{sec:typecheck}. A first observation is that it is
possible to build the automaton lazily, starting from the initial
states, producing new states and computing $\Phi(\_)$ only on
demand. This is sometimes useful since the emptiness check algorithm
we are going to describe in the next section works in a top-down way
and will not always materialize the whole automaton.

The defining equations for the function $\Inf$ as given in
Section~\ref{sec:typecheck} produce huge formulas. We will now
describe new equations that produce much smaller formulas in practice.
Before describing them, it is convenient to generalize the notation
$\Inf(e, q, \vec{q})$ by allowing a
{\em set of states} $\qbar \subseteq Q$ instead of a single state $q \in
Q$ for the output type. Intuitively, we want 
$\Inf(e, \qbar, \vec{q})$ to be semantically equivalent to
$\bigvee_{q \in \qbar} \Inf(e, q, \vec{q})$. We obtain a direct
definition of $\Inf(e,\qbar,\vec{q})$ by adapting the rules
for $\Inf(e,q,\vec{q})$:
$$
\begin{array}{lll}
\Inf(\an{b}{m}(e_1, \ldots, e_m), \qbar, \vec{q}) & = &
\displaystyle
\bigvee_{(q \from \an{b}{m}(\vec{q'}))\in \Delta, q \in \qbar}
\bigwedge_{j=1\ldots,m} \Inf(e_j, \{q_j'\}, \vec{q})
\\
\Inf(\an{p}{l}(x_h, e_1, \ldots, e_l), \qbar, \vec{q}) & = &
\displaystyle
  \bigvee_{\vec{q'} \in Q^l}
    \left(
    \down_h \tup{\an{p}{l}, \qbar, \vec{q'}} \wedge
    \bigwedge_{j=1,\ldots,l} \Inf(e_j, \{q_j'\}, \vec{q})
    \right)
\\
\Inf(y_j, \qbar, \vec{q}) & = &
\left\{
  \begin{array}{ll}
    \top & \hfill (q_j \in \qbar) \\
    \bot & \hfill (q_j \not \in \qbar) 
  \end{array}
\right.
\end{array}
$$ 
We have used the notation $\down_h \tup{\an{p}{l}, \qbar,
\vec{q'}}$.  Intuitively, this should be semantically equivalent to the
union $\bigvee_{q \in \qbar} \down_h \tup{\an{p}{l}, q, \vec{q'} }$.
Instead of using this as a definition, we prefer to change the set of
states of the automaton:
$$
\begin{array}{lll}
\Xi & = &
\{ \tup{\an{p}{k},\qbar,q_1,\ldots,q_k} \mid \an{p}{k}\in P, \; 
\qbar \subseteq Q, \vec{q} \in Q^k \}
\\
\Xi_0 & = & \{ \tup{p_0, Q_F} \mid p_0\in P_0 \}
\\
\Phi(\tup{\an{p}{k}, \qbar, \vec{q}},\an{a}{n}) & = & 
\bigvee_{(\an{p}{k}(\an{a}{n}(\vec{x}), \vec{y})=e) \in R} 
\Inf(e, \qbar, \vec{q}).
\end{array}
$$
In theory, this new alternating tree automaton could have
exponentially many more states. However, in practice, and because
of the optimizations we will describe now, this actually reduces
significantly the number of states that need to be computed.

The sections below will use the semantical equivalence $\bigvee_{q \in
  \qbar} \Inf(e, \{q\}, \vec{q}) \simeq \Inf(e, \qbar, \vec{q})$
mentioned above in order to simplify formulas.

\subsubsection{Cartesian factorization}

The rule for the constructor expression $\an{b}{m}(e_1,
\ldots, e_m)$ can be written:

$$
\Inf(\an{b}{m}(e_1, \ldots, e_m), \qbar, \vec{q}) =
\bigvee_{\vec{q'} \in \Delta(\qbar,\an{b}{m})}
\bigwedge_{j=1\ldots,m} \Inf(e_j, \{q_j'\}, \vec{q})
$$
where $\Delta(\qbar,\an{b}{m}) = \{ \vec{q'} ~|~ q \from
\an{b}{m}(\vec{q'})\in \Delta, q \in \qbar\} \subseteq Q^m$.
Now assume that we have a decomposition of this set
$\Delta(\qbar,\an{b}{m})$ as a union of $l$ Cartesian products:
$$\Delta(\qbar,\an{b}{m}) = (\qbar^1_1 \times \ldots \times
\qbar^1_m) \cup \ldots \cup (\qbar^l_1 \times \ldots \times
\qbar^l_m)$$
where the
$\qbar^i_j$ are sets of states. It is always possible to find
such a decomposition: at worst, using only singletons for the
$\qbar^i_j$, we will have as many terms in the union as $m$-tuples in
$\Delta(\qbar,\an{b}{m})$. But often, we can produce a decomposition with
fewer terms in the union. Let us write $\cart(\Delta(\qbar,\an{b}{m})$ for
such a decomposition (seen as a subset of $(2^Q)^m$). One can then
use the following rule:
$$
\Inf(\an{b}{m}(e_1, \ldots, e_m), \qbar, \vec{q}) =
\bigvee_{(\qbar_1,\ldots,\qbar_m) \in \cart(\Delta(\qbar,\an{b}{m}))}
\bigwedge_{j=1,..,m} \Inf(e_j, \qbar_j, \vec{q})
$$

\subsubsection{State partitioning}

\paragraph{Intuition}
The rule for procedure call enumerates all the possible states for the
value of parameters of the called procedure. In its current form,
this rule always produces a big union with $|Q|^l$ terms. However, it
may be the case that we don't need fully precise information about
the value of a parameter to do the backward type inference.  

Let us illustrate that with a simple example. Assume that the called
procedure $\an{p}{1}$ has a single parameter $y_1$ and that it never
does anything else with $y_1$ than copying it (that is, any rule for
$p$ whose right-hand side mentions $y_1$ is of the form
$\an{p}{1}(\an{a}{n}(x_1,\ldots,x_n),y_1) = y_1$). Clearly, all the
states $\tup{p, \qbar, q_1'}$ with $q_1' \in \qbar$ are equivalent,
and similarly for all the states $\tup{p, \qbar, q_1''}$ with $q_1''
\not \in \qbar$. This is because whether the result of the procedure
call will be or not in $\qbar$ only depends on the input tree (because
there might be other rules whose right-hand side don't involve $y_1$
at all) and on whether the value for the parameter is itself in
$\qbar$ or not. In particular, we don't know to know exactly in which
state the accumulator is.  So the rule for calling this procedure
could just be:

$$
\begin{array}{l}
\Inf(p(x_h, e_1), \qbar, \vec{q})
\\
~~~ = 
\displaystyle
  \bigvee_{q'_1 \in Q}
    \down_h \tup{p, \qbar, q'_1} \wedge
    \Inf(e_1, \{q'_1\}, \vec{q})
\\
~~~ =
\displaystyle
  \left(
  \bigvee_{q'_1 \in \qbar}
    \down_h \tup{p, \qbar, q'_1} \wedge
    \Inf(e_1, \{q'_1\}, \vec{q})
  \right) 
  \cup
  \left(
  \bigvee_{q''_1 \in Q \backslash \qbar}
    \down_h \tup{p, \qbar, q''_1} \wedge
    \Inf(e_1, \{q''_1\}, \vec{q})
  \right) 
\\
~~~ =
\displaystyle
    \left(
    \down_h \tup{p, \qbar, q_1'}
    \wedge
    \Inf(e_1, \qbar, \vec{q})
    \right)
    \vee
    \left(
    \down_h \tup{p, \qbar, q_1''}
    \wedge
    \Inf(e_1, Q \backslash \qbar, \vec{q})
    \right)
\end{array}
$$
where in the last line $q_1'$ (resp. $q_1''$) is chosen arbitrarily in
$\qbar$ (resp.  $Q \backslash \qbar$).

\paragraph{A new rule}
More generally, in the rule for a call to a procedure $\an{p}{l}$, we
don't need to consider all the $l$-tuples $\vec{q'}$, but only a subset
of them that capture all the possible situations.  First, we assume
that for given procedure $\an{p}{l}$ and output type $\qbar$,
one can compute for each $j=1,..,l$ an equivalence
relation $E\tup{\an{p}{l},\qbar,j}$ such that:
$$(\forall j=1,..,l.~(q'_j,q''_j) \in E\tup{\an{p}{l},\qbar,j})
\Rightarrow
\tup{\an{p}{l},\qbar,\vec{q'}} \simeq
\tup{\an{p}{l},\qbar,\vec{q''}}~~~(*)$$
Let
us look again at the right-hand side of the definition 
for $\Inf(\an{p}{l}(x_h,e_1,\ldots,e_l),\qbar,\vec{q})$:
$$
\Inf(\an{p}{l}(x_h, e_1, \ldots, e_l), \qbar, \vec{q}) =
\displaystyle
  \bigvee_{\vec{q'} \in Q^l}
    \left(
    \down_h \tup{\an{p}{l}, \qbar, \vec{q'}} \wedge
    \bigwedge_{j=1,\ldots,l} \Inf(e_j, \{q_j'\}, \vec{q})
    \right)
$$
Let us split this union according to the equivalence class
of the $q'_j$ modulo the relations  $E\tup{\an{p}{l},\qbar,j}$.
If for each $j$, we choose an equivalence class $\qbar_j$
for the relation $E\tup{\an{p}{l},\qbar,j}$ (we write $\qbar_j
\triangleleft E\tup{\an{p}{l},\qbar,j}$), then all the states
$\tup{\an{p}{l},\qbar,\vec{q'}}$ with $\vec{q'} \in \qbar_1 \times
\ldots \times \qbar_l$ are equivalent to
$\tup{\an{p}{l},\qbar,\Choice(\qbar_1 \times \ldots \times \qbar_l)}$,
where $\Choice$ is a choice function (it picks an arbitrary element
from its argument).  We can thus rewrite the right hand-side to:
$$
  \bigvee_{\qbar_1 \triangleleft E\tup{\an{p}{l},\qbar,1}, \ldots,
           \qbar_l \triangleleft E\tup{\an{p}{l},\qbar,l}}
    \left(
    \down_h \tup{\an{p}{l}, \qbar, \Choice(\qbar_1 \times \ldots
      \times \qbar_l)} \wedge
    \bigvee_{\vec{q'} \in \qbar_1 \times \ldots \times \qbar_l}
    \bigwedge_{j=1,\ldots,l} \Inf(e_j, \{q'_j\}, \vec{q})
    \right)
$$
The union of all the
formulas $\bigwedge_{j=1,..,l} \Inf(e_j,\{q'_j\},\vec{q})$ for
$\vec{q'} \in \qbar_1 \times \ldots \times \qbar_l$ is equivalent to
\break $\bigwedge_{j=1,..,l} \Inf(e_j,\qbar_j,\vec{q})$. 
Consequently, we obtain the following new rule:
$$
\begin{array}{l}
\Inf(\an{p}{l}(x_h, e_1, \ldots, e_l), \qbar, \vec{q}) = \\
~~~
\displaystyle
  \bigvee_{\qbar_1 \triangleleft E\tup{\an{p}{l},\qbar,1}, \ldots,
           \qbar_l \triangleleft E\tup{\an{p}{l},\qbar,l}}
    \left(
    \down_h \tup{\an{p}{l}, \qbar, \Choice(\qbar_1 \times \ldots
      \times \qbar_l)} \wedge
    \bigwedge_{j=1,\ldots,l} \Inf(e_j, \qbar_j, \vec{q})
    \right)
\end{array}
$$
In the worst case, all the equivalence relations
$E\tup{\an{p}{l},\qbar,j}$ are the identity, and
the right-hand side is the same as for the old rule. But if we can identify
larger equivalence classes, we can significantly reduce the number of terms
in the union on the right-hand side.

\paragraph{Computing the equivalence relations}
Now we will give an algorithm to compute the relations
$E\tup{\an{p}{k},\qbar,j}$ satisfying the condition $(*)$. 
We will also define equivalence relations $E[e,\qbar,j]$ for
any $(n,k)$-expression $e$ (with $j=1,..,k$), such that:
$$
(\forall j=1,..,k. (q'_j,q''_j) \in E[e,\qbar,j])
\Rightarrow
\Inf(e,\qbar,\vec{q'}) \simeq \Inf(e,\qbar,\vec{q''})
$$
We can use the rules used to define
the formulas $\Inf(e,\qbar,\vec{q})$ in order to obtain
sufficient conditions to be satisfied so that these properties hold.
We will express these conditions by a
 system of equations.  Before giving
 this system, we need to introduce some notations.
 If $E_1$ and $E_2$
 are two equivalence relations on $Q$, we write $E_1 \sqsubseteq E_2$
 if $E_2 \subseteq E_1$ (when equivalence relations are seen as
 subsets of $Q^2$). The smallest equivalence relation for this
 ordering is the equivalence relation with a single equivalence class.
 The largest equivalence relation is the identity on $Q$.  For two
 equivalence relations $E_1,E_2$, we can define their least upper
 bound $E_1 \sqcup E_2$ as the set-theoretic intersection. For an
 equivalence relation $E$ and a set of states $\qbar$, we write $\qbar
 \triangleleft E$ if $\qbar$ is one of the equivalence class modulo
 $E$. Abusing the notation by identifying an equivalence relation
 with the partition it induces on $Q$, we will write $\{Q\}$ for
 the smallest relation and $\{\qbar,Q \backslash \qbar\}$ for
 the relation with the two equivalence classes $\qbar$ and its complement.
 The system of equations is derived from the rules used to define
 the function $\Inf$:

$$
\begin{array}{lll}
E[\an{b}{m}(e_1, \ldots, e_m), \qbar, i] & \sqsupseteq &
\displaystyle
\bigsqcup \{ E[e_j,\qbar_j,i] ~|~
(\qbar_1,\ldots,\qbar_m) \in \cart(\Delta(\qbar,\an{b}{m})),~
j=1..m \} \\
E[\an{p}{l}(x_h, e_1, \ldots, e_l), \qbar, i] & \sqsupseteq &
\displaystyle
  \bigsqcup \{ E[e_j,\qbar_j,i] ~|~
    \qbar_j \triangleleft E\tup{\an{p}{l},\qbar,j},~
    j=1..l \} \\
E[y_j,\qbar,i] & \sqsupseteq &
\left\{
  \begin{array}{ll}
    \{\qbar,Q \backslash \qbar\} & \hfill (i = j) \\
    \{Q\} & \hfill (i \neq j) 
  \end{array}
\right. \\
E\tup{\an{p}{k},\qbar,j} & \sqsupseteq &
\displaystyle
 \bigsqcup \{ E[e,\qbar,j ] ~|~
 \an{p}{k}(\an{a}{n}(\vec{x}), \vec{y})=e) \in R \}
\end{array}
$$ 
Let us explain why these conditions imply the required properties for
the equivalence relation and how they are derived from the rules
defining $\Inf$. We will use an intuitive induction argument (on
expressions), even though a formal proof actually requires an
induction on trees.  Consider the rule for the procedure call. The new
rule we have obtained above implies that in order to have
$\Inf(\an{p}{l}(x_h,e_1,\ldots,e_l),\qbar,\vec{q'}) \simeq
\Inf(\an{p}{l}(x_h,e_1,\ldots,e_l),\qbar,\vec{q''})$, it is sufficient
to have $\Inf(e_j,\qbar_j,\vec{q'}) \simeq
\Inf(e_j,\qbar_j,\vec{q''})$ for all $j=1,..,l$ and for all $\qbar_j
\triangleleft E\tup{\an{p}{l},\qbar,j}$, and thus, by induction, it is
also sufficient to have $(q'_i,q''_i) \in E[e_j,\qbar_j,i]$ for all
$i$, for all $j=1,..,l$ and for all $\qbar_j \triangleleft
E\tup{\an{p}{l},\qbar,j}$.  In other words, a sufficient condition is
$(q'_i,q''_i) \in \bigcap \{ E[e_j,\qbar_j,i] ~|~ \qbar_j
\triangleleft E\tup{\an{p}{l},\qbar,j},~ j=1..l \}$, from which we
obtain the equation above (we recall that $\sqcup$ corresponds
to set-theoretic intersection of relations). The reasoning is similar
for the constructor expression. Indeed, the rule we have obtained in the
previous section tells us that in order to have
$\Inf(\an{b}{m}(e_1,\ldots,e_m),\qbar,\vec{q'}) \simeq
\Inf(\an{b}{m}(e_1,\ldots,e_m),\qbar,\vec{q''})$, it is sufficient
to have $\Inf(e_j,\qbar_j,\vec{q'}) \simeq
\Inf(e_j,\qbar_j,\vec{q''})$ for all $(\qbar_1,\ldots,\qbar_m)
\in \cart(\Delta(\qbar,\an{b}{m}))$ and $j=1,..,m$.

As we explained before, it is desirable to compute equivalence
relations with large equivalence classes (that is, small for the
$\sqsubseteq$ ordering). Here is how we can compute a family of
equivalence relations satisfying the system of equations above. First,
we consider the CPO of functions mapping a triple $(e,\qbar,i)$ to an
equivalence relation on $Q$ and we reformulate the system of equation
as finding an element $x$ of this CPO such that $f(x) \sqsubseteq x$,
where $f$ is obtained from the right-hand sides of the equations.
To compute such an element, we start from $x_0$ the smallest element
of the CPO, and we consider the sequence defined by $x_{n+1} =
x_n \sqcup f(x_n)$. Since this sequence is monotonic and the CPO
is finite, the sequence reaches a constant value after a finite number
of iterations. This value $x$ satisfies $f(x) \sqsubseteq x$ as
expected. We conjecture that this element is actually a smallest
fixpoint for $f$, but we have no proof of this fact (note that
the function $f$ is not monotonic).

\subsubsection{Sharing the computation}

Given the rules defining the formulas $\Inf(e,\qbar,\vec{q})$, we
might end up computing the same formula several times. A very
classical optimization consists in memoizing the results of such
computations. This is made even more effective by hash-consing
the expressions. Indeed, in practice, for a given mtt procedure,
many constructors have identical expressions.

\subsubsection{Complementing the output}
\label{sect:complementing}

In the example at the beginning of the previous subsection, we have
displayed a formula where both $\Inf(e,\qbar,\vec{q})$
and $\Inf(e,Q \backslash \qbar,\vec{q})$ appear. One may wonder
what is the relation between these two sub-formulas. Let us recall
the required properties for these two formulas:
$$\den{\Inf(e,\qbar,\vec{q})} = \{ v ~|~ \den{p}(\vec{v},\vec{w})
\cap \den{\qbar} \neq \emp \}$$
$$\den{\Inf(e,Q \backslash \qbar,\vec{q})} = \{ v ~|~ \den{p}(\vec{v},\vec{w})
\cap \den{Q \backslash \qbar} \neq \emp \}$$
(for $\vec{w} \in \den{\vec{q}}$). Note that $\den{Q
  \backslash \qbar}$ is the complement of $\den{\qbar}$.
As a consequence, if $\den{p}$ is a total deterministic function (that is,
if $\den{p}(\vec{v},\vec{w})$ is always a singleton), then
$\den{\Inf(e,Q \backslash \qbar,\vec{q})}$ is the complement
of $\den{\Inf(e,\qbar,\vec{q})}$. If we extend the syntax of formula
in alternating tree automata with negation (whose semantics is trivial
to define), we can thus introduce the following rule:
$$\Inf(e,\qbar,\vec{q}) = \neg \Inf(e,Q \backslash \qbar,\vec{q})$$
to be applied e.g. when the cardinal of $\qbar$ is strictly larger
than half the cardinal of $Q$. In practice, we observed a huge impact
of this optimization: the number of constructed states is divided by
two in all our experiences, and the emptiness algorithm runs much more
efficiently. Also, because of the memoization technique mentioned
above, this optimization allows us to share more computation. That
said, we don't have a clear explanation for the very important impact
of this optimization.

The rule above can only be applied when the expression $e$ denotes
a total and deterministic function. We use a very simple syntactic criterion
to ensure that: we require all the reachable procedures $\an{p}{k}$ to have
exactly one rule $\an{p}{k}(\an{a}{n}(x_1, \ldots, x_n),
y_1,\ldots,y_k) \to e$ for each symbol $\an{a}{n}$.

\subsection{Emptiness algorithm}
\label{sec:emptiness}

In this section, we describe an efficient algorithm to check emptiness
of an alternating tree automaton.
Instead of giving directly the final version of the algorithm which
would look quite obscure, we prefer to start describing formally a
simple algorithm and then explain various optimizations.

Let $\A = (\Xi,\Xi_0,\Phi)$ be an ata as defined in
Section~\ref{sec:ta}. Negation (as introduced in
Section~\ref{sect:complementing}) will be considered later when
describing optimizations.  The basic algorithm relies on a powerset
construction to translate $\A$ into a bottom-up tree automaton $\M =
(Q,Q_F,\Delta)$. We define $Q$ as the powerset $2^\Xi$. Intuitively, a
state $\Xbar = \{X_1,\ldots,X_m\}$ in $Q$ represents the intersection
of the ata states $X_i$. For such a state and a tag $\an{a}{n}$, one
must thus consider the formula $\varphi(\Xbar,\an{a}{n}) =
\bigwedge_{i=1,..,m} \Phi(X_i,\an{a}{n})$, and put in $\Delta$
transitions of the form $\Xbar \from
\an{a}{n}(\Xbar_1,\ldots,\Xbar_n)$ to mimic the formula
$\varphi(\Xbar,\an{a}{n})$.  First, we put $\varphi(\Xbar,\an{a}{n})$
in disjunctive normal form, using the $\DNF$ function introduced in
Section~\ref{sec:prelim}:

$$\varphi(\Xbar,\an{a}{n}) \simeq \bigvee_{(\Xbar_1,\ldots,\Xbar_n) \in \DNF(\varphi(\Xbar,\an{a}{n}))} 
  \bigwedge_{i=1,..,n} \bigwedge_{X \in \Xbar_i} \down_i X$$

The transition relation $\Delta$ consists of all
the transitions $\Xbar \from \an{a}{n}(\Xbar_1,\ldots,\Xbar_n)$
such that $(\Xbar_1,\ldots,\Xbar_n) \in \DNF(\varphi(\Xbar,\an{a}{n}))$.
One defines $Q_F = \{ \{ X \} ~|~ X \in \Xi_0 \}$.
One can easily establish that $\den{\Xbar}_{\M} = \bigcap_{X \in \Xbar}
\den{X}_{\A}$ and thus that $\lang(\M) = \lang(\A)$.

It is well-known that deciding emptiness of a bottom-up tree automaton
can be done in linear time. The classical algorithm to do so works in
a bottom up way and thus requires to fully materialize the automaton
(which is of exponential size compared to the original ata). However,
the construction above produces the automaton in a top-down way: for a
given state $\Xbar$, the construction gives all the transitions of the
form $\Xbar \from \ldots$. We can exploit this fact to derive an
algorithm that doesn't necessarily require the whole automaton $\M$ to
be built. The algorithm is given below in pseudo-code. The
function $\tt empty$ takes a state $\Xbar$ and returns $\tt true$ if
it is empty or $\tt false$ otherwise. The test is done under a number
of assertions represented by two global variables $\tt P$,$\tt N$
which stores sets of $\M$-states.  The set stored in $\tt P$
(resp. $\tt N$) represents positive (resp. negative) emptiness
assumptions: states which are assumed to be empty
(resp. non-empty). When the state $\Xbar$ under consideration is
neither in $\tt P$ or $\tt N$, it is first assumed to be empty (added
to $\tt P$). This assumption is then checked recursively by exploring
all the incoming transitions (for all possible tags and all components
of the disjunctive normal form corresponding to this tag) and if a
contradiction is found, the set of positive assumptions is backtracked
and $\Xbar$ is added to the set of negative assumptions.
This memoization-based scheme is standard for
coinductive algorithms.

\begin{alltt}
function empty (\(\Xbar\))
  if \(\Xbar \in\) P then return true
  if \(\Xbar \in\) N then return false
  let P_saved = P in
  P \(\leftarrow\) P \(\cup \{\Xbar\}\); 
  foreach \(\an{a}{n}\in\Sigma\) 
    if not (empty_formula (\(\varphi(\Xbar,\an{a}{n})\))) then
      P \(\leftarrow\) P_saved
      N := N \(\cup \{\Xbar\}\)
      return false
  return true

function empty_formula (\(\phi\))
  foreach \((\Xbar\sb{1},\ldots,\Xbar\sb{n})\in{}\DNF(\phi)\)
    if not (empty_sub \((\Xbar\sb{1},\ldots,\Xbar\sb{n})\)) then
      return false
  return true

function empty_sub \((\Xbar\sb{1},\ldots,\Xbar\sb{n})\) 
  foreach \(1\leq{}i{}\leq{}n\)
    if (empty \(\Xbar\sb{i}\)) then
      return true
  return false
\end{alltt}

This algorithm is not linear in the size of the automaton
$\M$ because of the backtracking on $\tt P$. This backtracking
can be avoided (as described in \cite{Frisch04PhD}, Chapter 7
or in \cite{SudaHosoya05}),
but the technique is rather
involved and would make the presentation of the optimizations
quite obscure. Moreover, we have indeed implemented the
non-backtracking version (with all the optimizations) but we did
not observe any noticeable speedup in our tests.

A first optimization improves the effectiveness of the memoization
sets $\tt P$ and $\tt N$. It is based on the fact that if $\Xbar_1 \subseteq
\Xbar_2$ then $\den{\Xbar_2} \subseteq \den{\Xbar_1}$. As a
consequence, if $\Xbar' \subseteq \Xbar$ for some $\Xbar' \in \tt P$,
then ${\tt empty}(\Xbar)$ can immediately return {\tt true}. Similarly,
if $\Xbar \subseteq \Xbar'$ for some $\Xbar' \in \tt N$, then ${\tt
empty}(\Xbar)$ can immediately return {\tt false}.

\paragraph{Enumeration and pruning of the disjunctive normal form}

The disjunctive normal form of a formula can be exponentially larger
than the formula itself. Our first improvement consists in not
materializing it but enumerating it lazily with a pruning technique
that avoids the exponential behavior in many cases.

\begin{alltt}
function empty_formula (\(\phi\))
  return (empty_dnf ([\(\phi\)],(\(\emptyset,\ldots,\emptyset)\)))

function empty_dnf (l,(\((\Xbar\sb{1},\ldots,\Xbar\sb{n})\) as a)) =
 match l with
 | [] -> return false
 | \(\top\) :: rest -> return (empty_dnf (rest,a))
 | \(\bot\) :: rest -> return true
 | \(\phi\sb{1}\vee\phi\sb{2}\) :: rest -> 
     if not (empty_dnf (\(\phi\sb{1}\) :: rest,a)) then return false
     return (empty_dnf (\(\phi\sb{2}\) :: rest,a))
 | \(\phi\sb{1}\wedge\phi\sb{2}\) :: rest -> 
     return (empty_dnf (\(\phi\sb{1}\)::\(\phi\sb{2}\)::rest,a))
 | \(\down\sb{h}X\) :: rest -> 
     if empty (\(\Xbar\sb{h}\cup\{X\}\))) then return true
     return (empty_dnf (rest,\((\Xbar\sb{1},\ldots,\Xbar\sb{h}\cup\{X\},\ldots,\Xbar\sb{n})\)))
\end{alltt}

The first argument of $\tt empty\_dnf$ is a list of formula
whose conjunction must be put in disjunctive normal form.
The second argument is an $n$-tuple (where $n$ is the arity of the
current symbol) which accumulates a ``prefix'' of the current term of
the disjunctive normal form being built. 
When an atomic formula $\down_h X$ is found, the state $X$ is added to
the $h$-th component of the accumulator. Here we have included an
important optimization: if the new state $\Xbar_h \cup \{X\}$ denotes
an empty set, then one can prune the enumeration. For instance, for a formula
of the form $\down_1 X \wedge \phi$ where $X$ turns out to be empty,
the enumeration will not even look at $\phi$.  This optimization
enforces the invariant that no component of the accumulator denotes
an empty set. As a consequence, when the function {\tt empty\_dnf}
reaches an empty list of formulas, the accumulator represents
an element of the disjunctive normal form for which {\tt empty\_sub}
would return {\tt false}.

The order in which we consider the two sub-formulas $\phi_1$
and $\phi_2$ in the formulas $\phi_1 \wedge \phi_2$ and $\phi_1 \vee
\phi_2$ might have a big impact on performances. It might be
worthwhile to look for heuristics guiding this choice.

\paragraph{Witness}

It is not difficult to see that the algorithm can be further
instrumented in order to produce a witness for non-emptiness (that is,
when ${\tt empty}(\Xbar)$ returns $\tt false$, it also returns a tree
$v$ which belongs to $\den{\Xbar}$). To do so, we keep for each state
in $\tt N$ a witness, and we also attach a witness to each component of
the accumulator $(\Xbar_1,\ldots,\Xbar_n)$ in the enumeration for the
disjunctive normal form. When checking for the emptiness of
$\Xbar\sb{h} \cup \{X\}$, we know that $\Xbar\sb{h}$ is a non-empty
state, and we have at our disposal a witness $v$ for this
state. Before doing the recursive call to $\tt empty$, we can first
check whether this witness $v$ is in $\den{X}$ (this can be done very
efficiently). If this is the case, we know that $\Xbar\sb{h} \cup
\{X\}$ is also non-empty. In practice, this optimization avoids many calls
to $\tt empty$.

\paragraph{Negation and reflexivity}

We have mentioned in Section~\ref{sect:complementing} an optimization
which introduces alternating formulas with negation. Using De
Morgan's laws, we can push the negation down and thus assume that it
can only appear immediately above an atomic formula $\down_i X$. Of
course, it is possible to get rid of the negation by introducing for
each state $X$ a dual state $\neg X$ whose transition formula (for
each tag) is the negation of the one for $X$; this only doubles the
number of states. However, we prefer to support directly in the
algorithm negated atomic formulas $\neg \down_i X$, because we can use
the very simple fact that it denotes a set which does not intersect
$\down_i X$. The algorithm is thus modified to work with pairs of sets
of $\A$-states, written $(\Xbar,\Ybar)$, which intuitively represents
the set $\bigcap_{X \in \Xbar} \den{X}_{\A} \backslash \bigcup_{Y
\in \Ybar} \den{Y}_{\A}$.  We define $\varphi((\Xbar,\Ybar),
\an{a}{n})$ as $\bigwedge_{X \in \Xbar} \Phi(X,\an{a}{n}) \wedge
\bigwedge_{Y \in \Ybar} \neg \Phi(Y,\an{a}{n})$.  The fact mentioned
above translates itself into a shortcut case in the $\tt empty$
function: if the input is $(\Xbar,\Ybar)$ with $\Xbar \cap \Ybar \not
= \emptyset$, then the result is {\tt true} (meaning that $(\Xbar,\Ybar)$
trivially denotes an empty set of trees).

The interesting cases for enumeration of the normal form are:
\begin{alltt}
 | \(\down\sb{h}X\) :: rest -> 
     if empty (\(\Xbar\sb{h}\cup\{X\}\))) then return true
     return (empty_dnf (rest,\(((\Xbar\sb{1},\Ybar\sb{1}),\ldots,(\Xbar\sb{h}\cup\{X\},\Ybar\sb{h}),\ldots,(\Xbar\sb{n},\Ybar\sb{n}))\)))
 | \(\neg\down\sb{h}Y\) :: rest -> 
     if empty (\(\Ybar\sb{h}\cup\{Y\}\))) then return true
     return (empty_dnf (rest,\(((\Xbar\sb{1},\Ybar\sb{1}),\ldots,(\Xbar\sb{h},\Ybar\sb{h}\cup\{Y\}),\ldots,(\Xbar\sb{n},\Ybar\sb{n}))\)))
\end{alltt}

\paragraph{Preprocessing}

Note the following trivial facts: For a formula $\phi_1 \wedge \phi_2$
to be empty, it is sufficient to have $\phi_1$ or $\phi_2$ empty; for
a formula $\phi_1 \vee \phi_2$ to be empty, it is sufficient to have
$\phi_1$ and $\phi_2$ empty; for a formula $\down_i X$ to be empty, it
is sufficient to have all the formulas $\Phi(X,\an{a}{n})$ empty; for
a formula $\neg \down_i X$ to be empty, it is sufficient to have all
the formulas $\neg \Phi(X,\an{a}{n})$ empty.

Using these sufficient conditions and a largest fixpoint computation,
we get a sound and efficient approximation of emptiness for formulas
(it returns $\tt true$ only if the formula is indeed empty, but it may
also return $\tt false$ is this case). We use this approximate
criterion to replace any subformula $\phi$ which is trivially empty
with $\bot$ and any subformula $\phi$ such that $\neg \phi$ is trivially
empty with $\top$ (and then apply Boolean tautologies to eliminate $\bot$
and $\top$ as arguments of $\vee$ or $\wedge$). In practice, this
optimization is very effective in reducing the size and complexity of
formulas involved in the real (exact) emptiness check.


\section{Experiments}
\label{sec:exper}

We have experimented on our typechecker with various XML
transformations implemented as mtts.  Although we did not try very big
transformations, we did work with large input and output tree automata
automatically generated from the XHTML
DTD (without taking XML attributes into account).  Note that because
this DTD has many tags, the mtts actually have many transitions since
they typically copy tags, which requires all constructors
corresponding to these tags to be enumerated.  They do not have too many
procedures, though.  The bottom-up deterministic automaton
that we generated from the XHTML DTD has 35 states.

Table~\ref{fig:experiment} gives the elapsed times spent in
typechecking several transformations and the number of states of the
inferred alternating tree automaton that have been materialized.  The
experiment was conducted on an Intel Pentium 4 processor 2.80Ghz,
running Linux kernel 2.4.27, and the typechecking time includes the
whole process (determinization of the output type, backward inference,
intersection with the input type, emptiness check). The typechecker is
implemented in and compiled by Objective Caml 3.09.3.

We also indicate the number of procedures in each mtt, the maximum
number of parameters, and the minimum integer $b$, if any, such that
the mtt is syntactically $b$-bounded copying. Intuitively, the integer
$b$ captures the maximum number of times the mtt traverses any node of
the input tree.  This notion has been introduced in \cite{Maneth07}
where the existence of $b$ is shown to imply the polynomiality of the
algorithm described in that paper (see also
Appendix~\ref{sec:maneth}). Here, we observe that even
unbounded-copying mtts can be typechecked efficiently.

\begin{table}[htbp]
  \centering
\begin{tabular}{|l|r|r|r|r|r|r|r|}
\hline
Transformation: & 
  (1) & (2) & (3) & (4) & (5) & (6) & (7)
\\
\hline
\# of procedures: & 2 & 2 & 3 & 5 & 4 & 6 & 6
\\
\hline
Max \# of parameters: & 1 & 1 & 1 & 1 & 2 & 2 & 2
\\
\hline
Bounded copying: & 1 & 1 & 2 & $\infty$ & $\infty$ & 2 & 1
\\
\hline\hline
Type-checking time (ms): &
  1057 &
  1042 &
  0373 &
  0377 &
  0337 &
  0409 &
  0410 
\\
\hline
\# of states in the ata: & 147 & 147 & 43 & 74 & 37 & 49 & 49
\\
\hline
\end{tabular}
\caption{Results of the experiments}
  \label{fig:experiment}
\end{table}

Unless otherwise stated, transformations are checked to have type
\verb|XHTML|$\to$\verb|XHTML| (i.e., both input and output types are
\verb/XHTML/).  Transformation (1) removes all the \verb|<b>| tags,
keeping their contents.  Transformation (2) is a variant that drops the
\verb|<div>| tags instead.  The typechecker detects that the latter
doesn't have type \verb|XHTML|$\to$\verb|XHTML| by producing a
counter-example:
$$
\verb|<html><head><title/></head><body><div/></body>|
$$
Indeed, removing the \verb|<div>| element may produce a
\verb|<body>| element with an empty content, which is not valid in
XHTML.
Transformation (3) copies all the \verb|<a>| elements (and their
corresponding subtrees) into a new \verb|<div>| element and prepends
the \verb/<div>/ to the \verb|<body>| element.  Transformation (4)
groups together adjacent \verb|<b>| elements, concatenating their
contents.
Transformation (5) extracts from an XHTML document a tree of depth 2
which represents the conceptual nesting structure of \verb|<h1>| and
\verb|<h2>| heading elements (note that, in XHTML, the structure among
headings is flat).  
Transformation (6) builds a tree representing a table of contents for
the top two levels of itemizations, giving section and subsection
numbers to them (where the numbers are constructed as Peano
numerals), and prepends the resulting tree to the \verb/<body>/
element.  Transformation (7) is a variant that only returns the table
of contents.


We have also translated some transformations (that can be expressed as
mtts) used by Tozawa and Hagiya in \cite{Tozawa04} (namely {\tt
  htmlcopy}, {\tt inventory}, {\tt pref2app}, {\tt pref2html}, {\tt
  prefcopy}). Our implementation takes between 2ms and 6ms to
typecheck these mtts, except for {\tt inventory} for which it takes 22
ms.  Tozawa and Hagiya report performance between 5ms and 1000ms on a
Pentium M 1.8 Ghz for the satisfiability check (which corresponds to
our emptiness check and excludes the time taken by backward
inference).  Although these results indicate our advantages over them
to some extent, since the numbers are too small and they have not
undertaken experiments as big as ours, it is hard to draw a meaningful
conclusion.


\section{Conclusion and Future Work}
\label{sec:concl}

We have presented an efficient typechecking algorithm for  mtts based
on the idea of using alternating tree automata for representing the
preimage of the given mtt obtained from the backward type inference.
This representation was useful for deriving optimization techniques on
the backward inference phase such as state partitioning and Cartesian
factorization, and was also effective for speeding up the subsequent
emptiness check phase by exploiting Boolean equivalences among
formulas.  Our experimental results confirmed that  our techniques
allow us to typecheck small sizes of transformations with respect to
the full XHTML type.  Finally, we have also made an exact connection
to two known algorithms, a classical one and Maneth-Perst-Seidl's, the
latter implying an important polynomial complexity under a
bounded-copying restriction.

The present work is only the first step toward a truly practical
typechecker for mtts.  In the future, we will seek for further
improvements that allow typechecking larger and more complicated
transformations.  In particular, transformations with upward axes can
be obtained by compositions of mtts as proved in \cite{Maneth05} and a
capability to typecheck such compositions of mtts in a reasonable time
will be important.  We have some preliminary ideas for the improvement
and plan to pursue them as a next step.  In the end, we hope to be
able to handle (at least a reasonably large subset of) XSLT.


\bibliographystyle{abbrv}
\bibliography{hh/haruo}

\appendix
\section{Comparison}
\label{sec:comparison}

In this section, we compare our algorithm with two existing
algorithms, the classical one based on function enumeration and the
Maneth-Perst-Seidl algorithm.

\subsection{Classical Algorithm}
\label{sec:classical}

The classical algorithm presented here is known as a folklore.
Variants can be found in the literature for deterministic mtts
\cite{Engelfriet03} and for macro forest transducers
\cite{Perst04}.  The algorithm takes a dbta $\M=(Q,Q_F,\Delta)$ and an mtt
$\T=(P, P_0, \Pi)$ and builds a dbta $\N' = (D, D_F, \delta)$ where:
$$
\begin{array}{lcl}
D & = & \{ \tup{\an{p}{m}, \vec{q}} \mid \an{p}{m}\in P,\;
\vec{q}\in Q^m \} \to 2^Q \\
D_F & = & \{ d \in D \mid p_0\in P_0,\; d(\tup{p_0}) \cap Q_F \neq \emp \} \\
\delta & = & \{ d \leftarrow \an{a}{n}(\vec{d}) \mid 
d(\tup{\an{p}{m}, \vec{q}}) = 
\bigcup_{(\an{p}{m}(\an{a}{n}(\vec{x}), \vec{y})\to e) \in \Pi}
\DInf(e, \vec{d}, \vec{q}) \}
\end{array}
$$
Here, the function $\DInf$ is defined as follows.
$$
\begin{array}{lcl}
\DInf(b^{(m)}(e_1, \ldots, e_m), \vec{d}, \vec{q}) & = &
\{ q' \mid
\begin{array}[t]{lll}
q' \leftarrow b^{(m)}(\vec{q'}) \in \Delta,\;
q_j' \in \DInf(e_j, \vec{d}, \vec{q}) ~~
\forall j=1,\ldots,m\;
\} 
\end{array}\\
\DInf(p(x_h, e_1,\ldots,e_l), \vec{d}, \vec{q}) & = &
\bigcup 
\{ d_h(\tup{p,\vec{q'}}) \mid q'_i \in \DInf(e_i, \vec{d}, \vec{q}), i=1,\ldots,l \} \\
\DInf(y_j, \vec{d}, \vec{q}) & = & \{ q_j \} 
\end{array}
$$
The constructed automaton $\N'$ has, as states, the set of all
functions that map each pair of a procedure and parameter types to a
set of states.  Intuitively, each state $d$ represents the set of
trees $v$ such that, given a procedure $\an{p}{m}$ and states
$\vec{q}$, the set of results of evaluating $p$ with the tree $v$ and
parameters $\vec{w}$ of types $\vec{q}$ is exactly described by
the states $d(\tup{p,\vec{q}})$.
Thus, the initial states $D_F$ represent the set of trees $v$ such
that the set of results from evaluating an initial procedure $p_0$
with $v$ contains a tree accepted by the given dbta $\M$.

The function $\DInf$ computes, from given expression $e$, states
$\vec{d}$ from $D$, and states $\vec{q}$ from $Q$, the set of states
that exactly describes the set of results of evaluating $e$ with a
tuple $\vec{v}$ of trees of types $\vec{d}$ and parameters of types
$\vec{q}$.   Then we can collect in $\delta$ transitions $d\from
\an{a}{n}(\vec{d})$ for all $\an{a}{n}$ and all $\vec{d}$ such that
$d$ is computed for all $\an{p}{m}$ and all $\vec{q}$ by using $\DInf$
with the expression on $\an{p}{m}$'s each rule for the symbol
$\an{a}{n}$.  By this intuition, each of the three cases for $\DInf$
can be understood as follows.
\begin{itemize}
\item The set of results of evaluating the constructor expression
  $b^{(m)}(e_1, \ldots, e_m)$ is described by the set of states $\vec{q'}$
  that have a transition $q' \leftarrow b^{(m)}(\vec{q'})
  \in \Delta$ such that each $q_i'$ describes the results of
  evaluating the corresponding subexpression $e_i$.
\item The set of results of evaluating the procedure call $p(x_h,
  e_1,\ldots,e_l)$ is the set of results of evaluating $p$ with the
  $h$-th input tree $v_h$ and parameters resulted from evaluating each
  $e_i$.  This set can be obtained by collecting the results of
  applying the function $d_h$ to $p$ and $\vec{q'}$ where each $q_i'$
  is one of the states that describe the set of results of $e_i$.
\item The set of results of evaluating the variable expression $y_j$ is
  exactly described by its type $q_j$.
\end{itemize}
Thus, the intuition behind is rather different from our approach.
Nevertheless, we can prove that the resulting automaton from the
classical algorithm is isomorphic to the one obtained from our
approach followed by determinization.

Determinization of an ata can be done as follows.  From an ata
$\A=(\Xi,\Xi_0,\Phi)$, we build a dbta $\N = (R, R_F, \Gamma)$
where
$$
\begin{array}{lcl}
R & = & 2^\Xi \\
R_F & = & \{ r \in \Xi \mid r \cap \Xi_0 \neq \emp \} \\
\Gamma & = &
\{ r \leftarrow \an{a}{n}(\vec{r}) \mid 
r = \{ X \mid \vec{r} \p \Phi(X,\an{a}{n}) \}
\}.
\end{array}
$$
Here, the judgment $\vec{r} \p \phi$ is defined inductively as follows.
\begin{itemize}
\item $\vec{r} \p \phi_1 \wedge \phi_2$ if $\vec{r} \p \phi_1$ and $\vec{r} \p \phi_2$.
\item $\vec{r} \p \phi_1 \vee \phi_2$ if $\vec{r} \p \phi_1$ or $\vec{r} \p \phi_2$.
\item $\vec{r} \p \top$.
\item $\vec{r} \p \down_i X$ if $X\in r_i$.
\end{itemize}
That is, $\vec{r} \p \phi$ intuitively means that $\phi$ holds by
interpreting each $\down_i X$ as ``$X$ is a member of the set $r_i$''.  

The intuition behind determinization of an ata is the same as that of
a nondeterministic tree automaton.  That is, each state $r$ in $\N$
denotes the set of trees $v$ that have type $X$ for all members $X$ of
$r$ and do not have type $Y$ for all non-members $Y$ of $r$.
\begin{equation}
  \label{eq:determinize0}
  \den{r} = \bigcap_{X\in r}\den{X} \setminus \bigcup_{Y\not\in r}\den{Y}
\end{equation}
This implies that any tree cannot have type $r$ and $r'$ at the same
time when $r\neq r'$.  Thus, the states of the tree automaton $\N$
form a partition of all the trees, that is, $\N$ is complete and
deterministic.  From this, we can understand the equivalence between
$\A$ and $\N$ since each final state in $\N$ contains an initial state
in the original ata $\A$ and therefore the set of such final states
forms a partition of the sets denoted by the initial states of $\A$.
Then, by using the formula (\ref{eq:determinize0}), the interpretation
``$X$ is contained in $r_i$'' of $\down_i X$ in the judgment
$\vec{r}\p \phi$ implies that $\den{r_i}\subseteq \den{X}$.  Here, we
can see a parallelism between the intuition of the judgment $\vec{v}\p
\phi$ (where $\down_i X$ is interpreted ``$v_i \in \den{X}$'') and
that of $\vec{r}\p \phi$.  Indeed, a key property to the proof below
is: $\vec{v}\p \phi$ if and only if $\vec{r}\p\phi$ for some $\vec{r}$
such that $\vec{v}\in\den{\vec{r}}$.

\begin{proposition}
  $\A$ and $\N$ are equivalent.
\end{proposition}

\begin{pf}
  To prove the result, it suffices to show the following.
  \begin{equation}
    \label{eq:determinize1}
    v \in \den{r}  \iff  r = \{ X \mid v \in \den{X} \}.
  \end{equation}
  (Note that this is a rewriting of the equation (\ref{eq:determinize0}).)
  Indeed, this implies
  \begin{eqnarray*}
    v\in \lang(\N) 
    & \iff &
    v\in \den{R_F} 
    \\
    & \stackrel{by \mathrm{(\ref{eq:determinize1})}}{\iff} &
    \exists r.~ (r \cap \Xi_0 \neq \emp \; \wedge\; r = \{ X \mid v \in \den{X} \})
    \\
    & \iff &
    \exists X \in \Xi_0.\; v\in \den{X}
    \\
    & \iff &
    v \in \lang(\A).
  \end{eqnarray*}
  The proof proceeds by induction on the structure of $v$.
  To show (\ref{eq:determinize1}), the following is sufficient
  \begin{equation}
    \label{eq:determinize2}
    (\exists\vec{r}.\;
    \vec{v} \in \den{\vec{r}} \;\;\wedge\;\; 
    \vec{r} \p \phi)
    \iff \vec{v} \p \phi.
  \end{equation}
  since this implies (\ref{eq:determinize1}):
  \begin{eqnarray*}
    \an{a}{n}(\vec{v}) \in \den{r} 
    & \iff &
    \exists (r \from \an{a}{n}(\vec{r}))\in \Gamma.\; \vec{v} \in \den{\vec{r}} 
    \\
    & \iff &
    \exists \vec{r}.\; r = \{ X \mid \vec{r} \p \Phi(X,\an{a}{n}) \}
    \;\wedge\; 
    \vec{v} \in \den{\vec{r}}
    \\
    & \stackrel{by \mathrm{(\ref{eq:determinize2})}}{\iff} &
    r = \{ X \mid \vec{v} \p \Phi(X,\an{a}{n}) \}
    \\
    & \iff &
    r = \{ X \mid \an{a}{n}(\vec{v}) \in \den{X} \}.
  \end{eqnarray*}
  The proof of (\ref{eq:determinize2}) itself is done by induction on
  the structure of $\phi$.  The ``only if'' direction is
  straightforward.  For the ``if'' direction,  let $r_i = \{ X \mid
  v_i \in \den{X} \}$ for $i=1,\ldots,n$.  By the \ih,
  (\ref{eq:determinize1}) gives $v_i \in \den{r_i}$.  The rest is case
  analysis on $\phi$.
  \begin{itemize}
  \item Case $\phi=\bot$.  This never arises.
  \item Case $\phi=\top$.  This case trivially holds.
  \item Case $\phi=\down_h X$.  From $\vec{v}\p\phi$, we have $v_h\in
    \den{X}$ and therefore $X\in r_h$ by the definition of $r_h$.
    This implies the result.
  \item Case $\phi=\phi_1\wedge\phi_2$.  By the \ih,  $\vec{v} \in
    \den{\vec{r'}}$ and $\vec{r'} \p \phi_1$ with $\vec{v} \in
    \den{\vec{r''}}$ and $\vec{r''} \p \phi_2$  for some $\vec{r'}$
    and $\vec{r''}$.  Since $\N$ is deterministic,  both $\vec{r'}$
    and $\vec{r''}$ actually equal to $\vec{r}$.  Hence the result
    follows.
  \item Case $\phi=\phi_1\vee\phi_2$.  Similar to the previous case.
    \endofpf
  \end{itemize}
\end{pf}

\begin{proposition}
  Let $\N$ be obtained by determinizing the ata from the last section.
  Then, $\N$ and $\N'$ are isomorphic.
\end{proposition}

\begin{pf}
Define the function $\beta$ from $D$ to $R$ as follows:
$$
\beta(d) =
\{ \tup{\an{p}{m}, q, \vec{q}} \mid 
\an{p}{m}\in P,\; \vec{q} \in Q^m, \; q\in d(\tup{p, \vec{q}}) \}
$$
Clearly, $\beta$ is bijective: $\beta^{-1}(r)(\tup{p, \vec{q}})
= \{ q \mid \tup{\an{p}{m}, q, \vec{q}}\in r \}$.  It remains
to show that $\beta$ is an isomorphism between $\N$ and $\N'$, that
is, (1) $\beta(D_F) = R_F$ and (2) $\beta(\delta(d)) = \Gamma(\beta(d))$ for
each $d$.  

The condition (1) clearly holds since $d(p_0)\cap Q_F \neq \emp$
iff $\tup{p_0,q} \in \beta(d)$ for some $q\in Q_F$.
To prove (2), it suffices to show 
\begin{quote}
  $q \in \DInf(e, \vec{d}, \vec{q})$
  iff $\beta(\vec{d}) \p \Inf(e, q, \vec{q})$.
\end{quote}
Here, $\beta(d_1,\ldots,d_k)$ stands for $(\beta(d_1),\ldots,\beta(d_k))$.
The proof is by induction on the structure of $e$.
\begin{itemize}
\item Case $e=b^{(m)}(e_1, \ldots, e_m)$.  
  \begin{eqnarray*}
    q \in \DInf(e, \vec{d}, \vec{q}) 
    & \iff & 
    \exists (q \leftarrow b^{(m)}(\vec{q'})) \in \Delta.\;
    \forall j.\; q_j' \in \DInf(e_j, \vec{d}, \vec{q}) \\
    & \stackrel{\mathrm{by I.H.}}{\iff} & 
    \exists (q \leftarrow b^{(m)}(\vec{q'})) \in \Delta.\;
    \forall j.\; 
    \beta(\vec{d}) \p \Inf(e_j, q_j', \vec{q}) \\
    & \iff &
    \beta(\vec{d}) \p
    \bigvee_{(q \from \an{b}{m}(\vec{q'}))\in \Delta}
    \bigwedge_{j=1\ldots,m} \Inf(e_j, q_j', \vec{q}) \\
    & \iff &
    \beta(\vec{d}) \p
    \Inf(e, q, \vec{q})
  \end{eqnarray*}
\item Case $e=p(x_h, e_1, \ldots, e_l)$. 
  \begin{eqnarray*}
    q \in \DInf(e, \vec{d}, \vec{q}) 
    & \iff & 
    \bigcup 
    \{ d_h(p,\vec{q'}) \mid q'_i \in \DInf(e_i, \vec{d}, \vec{q}), i=1,\ldots,l \} \\
    & \iff &
    \exists \vec{q'}. \;
    q \in d_h(p,\vec{q'}) \mbox{ and }
    \forall i.\; q_i' \in \DInf(e_i, \vec{d}, \vec{q'}) \\
    & \stackrel{\mathrm{by I.H.}}{\iff} & 
    \exists \vec{q'}. \;
    \tup{p,q,\vec{q'}} \in \beta(d_h) \mbox{ and }
    \forall i.\; \beta(\vec{d}) \p \Inf(e_i, q, \vec{q'}) \\
    & \iff &
    \beta(\vec{d}) \p
    \bigvee_{\vec{q'} \in Q^l} 
    \bigwedge_{i=1,\ldots,l}
    \Inf(e_i,q,\vec{q'})
    \wedge
    \down_i \tup{p, q, \vec{q'}} \\
    & \iff &
    \beta(\vec{d}) \p
    \Inf(e, q, \vec{q})
  \end{eqnarray*}
\item Case $e=y_j$.  First, 
  $q \in \DInf(y_j, \vec{d}, \vec{q})$ iff
  $q = q_j$.  If $q=q_j$, then $\Inf(e, q, \vec{q})=\top$ and
  therefore the RHS holds.  If $q\neq q_j$, then 
  $\Inf(e, q, \vec{q})=\bot$ and therefore the RHS does not hold.
  \endofpf
\end{itemize}
\end{pf}

\subsection{Maneth-Perst-Seidl Algorithm}
\label{sec:maneth}

First, for simplicity in comparing the two algorithms, following
\cite{Maneth07}, we consider an mtt where the input type is
already encoded into procedures.  That is, instead of the original mtt
$\T$, we take an mtt $\T'$ and a bta $\M_\IN$ such that
$$
\T'(v) = 
\left\{
\begin{array}{ll}
\T(v) & (v \in \lang(\M_\IN)) \\
\emp & (\mbox{otherwise}).
\end{array}
\right.
$$
That is, $\T'$ behaves exactly the same as $\T$ for the inputs from
$\lang(\M_\IN)$ but returns no result for the other inputs.  See
\cite{Maneth07} for a concrete construction.  Having done
this, we only need to check that $\{ v \mid \T'(v)\cap
\lang(\M) \neq \emp \} = \emp$.

In Maneth-Perst-Seidl algorithm, we construct a new mtt $\U$ from
$\T'=(P,P_0,\Pi)$ specialized to the output-type dbta $\M=(Q, Q_F,
\Delta)$ such that $\U(v) = \T'(v)\cap \lang(\M)$ for any tree $v$.
This can be done by constructing the mtt $\U=(S,S_0,\Omega)$ where
$$
\begin{array}{lll}
  S & = & \{ \an{\tup{\an{p}{m}, q, \vec{q}}}{m} \mid \an{p}{m} \in P,
  q, \vec{q} \in Q^m \} \\
  S_0 & = & \{ \tup{p_0,q} \mid p_0\in P_0,\; q\in Q_F \} \\
  \Omega & = & 
  \{ \tup{\an{p}{m}, q, \vec{q}}(\an{a}{n}(\vec{x}),\vec{y}) \to e' \mid
  (\an{p}{m}(\an{a}{n}(\vec{x}),\vec{y}) \to e) \in \Pi,\;
  e'\in \Spec(e,q,\vec{q}) \}.
\end{array}
$$
Here, we define the function $\Spec$ as follows.
$$
\begin{array}{lll}
  \Spec(a(e_1,\ldots,e_n),q,\vec{q}) & = &
  \{ a(e_1',\ldots,e_n')\mid
  q \leftarrow a(q_1',\ldots,q_n')\in\Delta,\;
  \forall i.\; e_i' \in \Spec(e_i, q_i', \vec{q}) \} 
  \\
  \Spec(p(x_h,e_1,\ldots,e_l),q,\vec{q}) & = &
  \{ \tup{p,q,\vec{q'}}(x_h,e_1',\ldots,e_l') \mid
  \vec{q'}\in Q^l,\; 
  \forall i.\; e_i' \in \Spec(e_i,q_i',\vec{q}) \} 
  \\
  \Spec(y_i,q,\vec{q}) & = & 
  \{y_i\} 
\end{array}
$$
Intuitively, each procedure $\tup{p, q, \vec{q}}$ in the new mtt
$\U$ yields, for any input value $v$ and for any parameters $\vec{w}$
of types $\vec{q}$, the same results as $p$ but restricted to type
$q$:
$$
\den{\tup{\an{p}{m}, q, \vec{q}}}(v, \vec{w}) =
\den{\an{p}{m}}(v, \vec{w}) \cap \den{q}
$$
Similarly, $\Spec(e,q,\vec{q})$ yields, for any input values
$\vec{v}$ and for all parameters $\vec{w}$ of types $\vec{q}$,  the
same results as $e$ but restricted to type $q$:
$$
\den{\Spec(e,q,\vec{q})}(\vec{v}, \vec{w}) =
\den{e}(\vec{v}, \vec{w}) \cap \den{q}
$$
After thus constructing the mtt $\U$, the remaining is to check
that the translation of $\U$ is empty, i.e., $\U(v)=\emp$ for any
value $v$.  This can be done as follows.  Define first the following
system of implications $\rho'$ where we introduce propositional
variables $\Xbar$ consisting of all subsets of $S$:
$$
\begin{array}{lll}
\rho' & = &
\{ 
\Xbar \Leftarrow \Xbar_1\wedge \ldots \wedge \Xbar_n
\mid
\begin{array}[t]{lll}
\exists \an{a}{n}.~
\exists e_1,\ldots,e_k.~
\forall \an{s}{m}\in \Xbar.~
\exists j.~
(\an{s}{m}(\an{a}{n}(\vec{x}),\vec{y})\to e_j)\in \Omega,\\
\forall i=1,\ldots,n.~
\Xbar_i = \{ s'\in S \mid \exists j=1,\ldots,k.~ s'(x_i,\ldots) \mbox{ occurs in } e_j \}
\}
\end{array}
\end{array}
$$
and then verify that $\rho'\p \{s\}$ for some $s\in S_0$.
Intuitively, each propositional variable $\Xbar$ denotes whether there
is  some input $v$ from which any procedure in the set $\Xbar$ translates to some
value with some parameters:
$$
\exists v.~ 
\forall \an{s}{m} \in \Xbar.~
\exists \vec{w}.~
\den{\an{s}{m}}(v,\vec{w}) \neq \emp
$$

Now, we can prove that the system of implications obtained from the MPS
and the one from our algorithm are exactly the same.  From this, we can
directly carry over useful properties found for the MPS algorithm to
our algorithm.  In particular, our algorithm has the same polynomial
time complexity under the restriction of a finitely bounded number of
copying \cite{Maneth07}.

\begin{proposition}
  Given an input type that accepts all trees and the mtt $\T'$ defined
  above, let $\A$ and $\rho$ be the ata and  the system of implications
  obtained by the algorithm in Section~\ref{sec:typecheck}.  Let
  $\Xi_0$ be $\A$'s initial states.  Then, $(\rho,\Xi_0)$ and
  $(\rho',S_0)$ are identical.
\end{proposition}

\begin{pf}
  Note that both $\rho$ and $\rho'$ consist of all variables $\Xbar$ where $\Xbar$ is from
  the set $P\times Q\times Q^m$.  The result follows by showing 
  $\Xbar\Leftarrow \Xbar_1\wedge\ldots\wedge \Xbar_n\in \rho$ iff
  $\Xbar\Leftarrow \Xbar_1\wedge\ldots\wedge \Xbar_n\in \rho'$.  It suffices to show
for any $\Xbar$ and $i$,  
$$
\exists e_1,\ldots,e_k.~
\forall s \in \Xbar.~
\exists j.~
(s(a(\vec{x}),\vec{y})\to e_j)\in \Omega,
\Xbar_i = \{ s'\in S \mid \exists j=1,\ldots,k.~ s'(x_i,\ldots) \mbox{ occurs in } e_j \}
$$
iff
$$
(\Xbar_1,\ldots,\Xbar_n) \in \DNF(\bigwedge_{s\in \Xbar}\Phi(s,a)).
$$
This follows by showing that, for all $(\Xbar_1,\ldots,\Xbar_n) \in
\DNF(\Inf(e_1,q_1,\vec{q}_1)\wedge\ldots\wedge\Inf(e_k,q_k,\vec{q}_k))$,
$$
\exists j=1,\ldots,k.~ s'(x_i) \mbox{ occurs in } \Spec(e_j,q_j,\vec{q}_j)
\iff s'\in \Xbar_i.
$$
This can be proved by induction on $|e_1|+\ldots+|e_k|$ where
$|e|$ is the size of $e$.%
\end{pf}

\begin{corollary}
  For any $b$-bounded copying mtt, our algorithm runs in polynomial time.
\end{corollary}

\section{Alternating tree automata with bounded traversing}
\label{sec:bounded}

The corollary in the last section depends on the proof of polynomiality
from~\cite{Maneth07}. It gives the information that the emptiness
check for alternating automata has polynomial time complexity
when the alternating automata is obtained by the basic backward
inference algorithm from Section~\ref{sec:typecheck} when applied
to a $b$-bounded copying mtt. It seems natural to look for a
counterpart of the notion of $b$-bounded copying for alternating
automata that directly ensures the polynomiality of the emptiness
check.

Let $\A = (\Xi,\Xi_0,\Phi)$ be an ata. For each state $X \in \Xi$,
we define the maximal traversal number $b[X]$ as the least fixpoint
of a constraint system over ${\cal N} = \{ 1 < 2 < \ldots < \infty
\}$, the complete lattice of naturals extended with $\infty$. The
constraint system consists of all the constraints of the form:
$$ b[X] \geq b_i[\Phi(X,\an{a}{n})] $$
for $\an{a}{n} \in \Sigma$ and $1 \leq i \leq n$, where
$b_i[\phi]$ is defined inductively:
$$
\begin{array}{lll}
b_i[\top] & = & 0 \\
b_i[\bot] & = & 0 \\
b_i[\phi_1 \wedge \phi_2] & = & b_i[\phi_1] + b_i[\phi_2] \\
b_i[\phi_1 \vee \phi_2] & = & \max(b_i[\phi_1],b_i[\phi_2]) \\
b_i[\down_h X] & = &
  \left\{
  \begin{array}{l}
  b[X] \mbox{~if~} i = h \\
  0 \mbox{~if~} i \not = h
  \end{array}
  \right.
\end{array}
$$
The ata $\A$ is (syntactically) $b$-bounded traversing if $b[X] \leq
b$ for all $X \in X_0$.

We mention without proving it formally that when we apply our backward
inference algorithm to a $b$-bounded copying mtt, then the resulting
ata is $b$-bounded traversing. More precisely, we can show
that $b[\tup{\an{p}{k},q,\vec{q}}] \leq b[\an{p}{k}]$ where
$b[\an{p}{k}]$ denotes the maximal copy number for the procedure
$\an{p}{k}$, as defined in \cite{Maneth07}. As a matter of fact,
the optimizations given in Section~\ref{sec:backwardalgo} preserve this
property (but the ata formally has exponentially many more states,
even if in practice only a fraction of them is going to be materialized).

Now it remains to establish that the emptiness check for a $b$-bounded
traversing ata runs in polynomial time. We define $b[\Xbar]$ as
$\Sigma_{X \in \Xbar} b[X]$. For any $b$-formula $\phi$ and
$(\Xbar_1,\ldots,\Xbar_n) \in \DNF(\phi)$ and $1 \leq i \leq n$, we
observe that $b[\Xbar_i] \leq b_i[\phi]$. The proof is by induction on
the structure of $\phi$. As a consequence, for any
$(\Xbar_1,\ldots,\Xbar_n) \in \DNF(\bigwedge_{X \in \Xbar}
\Phi(X,\an{a}{n}))$, we have $b[\Xbar_i] \leq b[\Xbar]$. So, if the
ata is $b$-bounded traversing, then the emptiness check algorithm will
only consider set of states $\Xbar$ such that $b[\Xbar] \leq b$. Since
$b[\Xbar]$ is a lower bound for the cardinal of $\Xbar$ (because $b[X]
\geq 1$ for all $X$), we see that the algorithm only looks at a
polynomial number of set of states $\Xbar$.

To conclude this section, we observe that the intersection of a
$b$-bounded traversal ata and a $b'$-bounded traversal ata is a
$(b+b')$-bounded traversal ata, and that a non-deterministic tree
automaton is isomorphic to a $1$-bounded traversal ata. This is
useful to typecheck a $b$-bounded copying mtt, because we need
to compute the intersection of the inferred ata, which is $b$-bounded
traversal, and of the input type, which is given by a
non-deterministic tree automaton. As a result, we obtain a
$(b+1)$-bounded ata.


\newpage
\tableofcontents

\end{document}